\documentclass[12pt]{article}
\usepackage{amsfonts,amssymb,amsmath,mathrsfs}
\usepackage{hyperref}
\newcommand{\beq}{\begin{equation}}
\newcommand{\eeq}{\end{equation}}
\newcommand{\vp}{\vphantom}

\newcommand{\wt}{\widetilde}
\newcommand{\de}{\delta}
\newcommand{\s}{\sigma}
\newcommand{\mc}{\mathcal}
\newcommand{\mr}{\mathrm} 
\newcommand{\supplus}{\lefteqn{\supset}\hspace{-0.1cm}+}
\begin{document}
\begin{center}
{\Large\bf Conformal higher-spin symmetries in twistor string theory}\\[0.5cm]
{\large D.V.~Uvarov\footnote{E-mail: d\_uvarov@\,hotmail.com}}\\[0.2cm]
{\it NSC Kharkov Institute of Physics and Technology,}\\ {\it 61108 Kharkov, Ukraine}\\[0.5cm]
\end{center}
\begin{abstract}
It is shown that similarly to massless superparticle, classical global symmetry of the Berkovits twistor string action is infinite-dimensional. We identify its superalgebra, whose finite-dimensional subalgebra contains $psl(4|4,\mathbb R)$ superalgebra. In quantum theory this infinite-dimensional symmetry breaks down to $SL(4|4,\mathbb R)$ one.
\end{abstract}

\setcounter{equation}{0}
\def\theequation{\thesection.\arabic{equation}}
\section{Introduction}
Twistor string theory \cite{Witten03}, \cite{Berkovits} inspired
remarkable progress in understanding spinor and twistor structures
underlying scattering amplitudes in gauge theories and gravity. Unlike conventional
superstrings the twistor string spectrum presumably includes only
a finite number of oscillation modes, in particular those of the
open string sector are exhausted by 4-dimensional $N=4$
superYang-Mills theory and conformal supergravity \cite{BerkWitt}.
Since the latter theory is non-unitary and one is unable beyond
the tree level to disentangle its modes from those of
superYang-Mills, there were made over time other propositions of
twistor string models \cite{Abou-Zeid}, \cite{Skinner},
\cite{MasonSkinner}. However, for tree-level gluon amplitudes
there was proved \cite{DolanGoddard} the equivalence of the
expressions obtained within the Berkovits model \cite{Berkovits}
and using the field-theoretic approach. To gain further insights
into the properties of twistor strings it is helpful to identify
their symmetries both classical and quantum. In \cite{Witten03},
\cite{Dolan} it was shown that except for an obvious $PSL(4|4,\mathbb R)$
global symmetry twistor strings are also invariant under its
Yangian extension that is closely related to infinite-dimensional
symmetry of integrable $N=4$ superYang-Mills theory
\cite{Zarembo}, \cite{Nappi}.

In this paper we argue that the world-sheet action of Berkovits
twistor string is invariant under infinite-dimensional global
symmetry. Its superalgebra contains finite-dimensional subalgebra spanned by the generators of $PSL(4|4,\mathbb R)$, 'twisted' $GL_t(1,\mathbb R)$ symmetries and constant shifts of supertwistor components.   
For the twistor string model with ungauged $gl(1,\mathbb R)$
current global symmetry algebra is isomorphic to the Dirac brackets (D.B.) algebra of the collection of all monomials constructed from an arbitrary number $L\geq0$ of $PSL(4|4,\mathbb R)$ supertwistors and the dual supertwistor. We identify this infinite-dimensional superalgebra as a twistor string algebra (TSA). Its finite-dimensional subalgebra is spanned by $gl(4|4,\mathbb R)$ generators and dual supertwistor corresponding to $L=1$ and $L=0$ monomials respectively. The relations of the global symmetry algebra of the Berkovits model are obtained from those of TSA by setting to zero $gl(1,\mathbb R)$
current. In the quantum theory
we show that classical inifinite-dimensional symmetry breaks down to
$SL(4|4,\mathbb R)$ one, whose consistency was proved in \cite{Dolan}
using the world-sheet CFT techniques.

Infinite-dimensional nature of the symmetries of massless superparticles was revealed already in \cite{Townsend}. So in Section 2 we consider the (higher-spin) symmetries of $N=4$ supersymmetric models of massless particles in the supertwistor formulation \cite{Shirafuji}, \cite{Claus}. We included in this section also some of the known material, in particular on the finite-dimensional symmetries and spectrum identification, to make it self-contained and to prepare the ground for subsequent discussion of the twistor string symmetries in Section 3.

\setcounter{equation}{0}
\section{Higher-spin symmetries of $D=4$ $N=4$ massless superparticles}

Kinetic term of the Shirafuji superparticle model \cite{Shirafuji} specialized to the case of $N=4$ supersymmetry \cite{Claus} has the form
\beq\label{sparticle}
S=\int d\tau\mathscr L,\quad\mathscr L=\frac{i}{2}(\bar{\mc Z}_A\dot{\mc Z}^A-\dot{\bar{\mc Z}}_A\mc Z^A).
\eeq
$PSU(2,2|4)$ supertwistor $\mc Z^A$ has 4 bosonic components $Z^\alpha$ transforming in the fundamental representation of $SU(2,2)\sim SO(2,4)$ and 4 fermionic components $\xi^i$ -- in the fundamental representation of $SU(4)$ \cite{Ferber}. Components of the dual supertwistor
\beq\label{dualstwistor}
\bar{\mc Z}_A=(\bar Z_\alpha,\bar\xi_i)=(\mc Z)^\dagger\mc H,\quad\mc H=\left(
\begin{array}{ccc}
0&I_{2\times2}& \\
I_{2\times2}&0& \\
& & I_{4\times4}
\end{array}
\right)
\eeq
transform in the antifundamental representation of $SU(2,2)\times SU(4)$.

In the canonical formulation non-trivial D.B. of the supertwistor components are
\beq\label{DirBr}
\{\mc Z^A,\bar{\mc Z}_B\}_{D.B.}=i\de^A_B,\quad\{\bar{\mc Z}_B,\mc Z^A\}_{D.B.}=-i(-)^a\de^A_B,
\eeq
where $a$ is the Grassmann parity equal 0 for the supertwistor bosonic components and 1 for the fermionic components.

\subsection{Classical symmetries of $D=4$ $N=4$ massless superparticle}

\subsubsection{$U(2,2|4)$ global symmetry}

Action (\ref{sparticle}) is manifestly invariant under $U(2,2|4)$ global symmetry generated by $\mathrm G_{(1,1)}=\bar{\mc Z}_B\Lambda^B\vp{\Lambda}_A\mc Z^A$: 
\beq\label{u224-stwistor-transformation}
\de\mc Z^A=\{\mc Z^A,\mr G_{(1,1)}\}_{D.B.}=i\Lambda^A\vp{\Lambda}_B\mc Z^B,\quad\de\bar{\mc Z}_A=\{\bar{\mc Z}_A,\mr G_{(1,1)}\}_{D.B.}=-i\bar{\mc Z}_B\Lambda^B\vp{\Lambda}_A.
\eeq
Supertwistor $\mc Z^A$ and its dual $\bar{\mc Z}_A$ are thus transform linearly under $U(2,2|4)$. Associated Noether current up to a numerical factor coincides with the generator of $U(2,2|4)$ transformations and is given by valence $(1,1)$ composite supertwistor
\beq\label{u224-stwistor'}
\mr T_A\vp{\mr T}^B=\bar{\mc Z}_A\mc Z^B
\eeq
that on D.B. satisfies the relations of $u(2,2|4)$ superalgebra
\beq\label{u224-algebra}
\{\mr T_A\vp{\mr T}^B,\mr T_C\vp{\mr T}^D\}_{D.B.}=i(\de^B_C\mr T_A\vp{\mr T}^D-(-)^{\varepsilon^b_a\varepsilon^d_c}\de^D_A\mr T_C\vp{\mr T}^B),\quad\varepsilon^b_a=(-)^{a+b}.
\eeq

Irreducible components of (\ref{u224-stwistor'})
\beq\label{u224-generators-comp}
\mr T_A\vp{\mr T}^B=\{\wt{\mr T}_\alpha\vp{\wt{\mr T}}^\beta,\ \wt{\mr T}_i\vp{\wt{\mr T}}^j;\ \mr Q_\alpha\vp{Q}^j,\ \mr Q_i\vp{Q}^\beta;\ \mr T,\ \mr U\}
\eeq
include the generators of $SU(2,2)\times SU(4)$ transformations, $D=4$ $N=4$ Poincare and conformal supersymmetries, $U(1)$ phase rotation and 'twisted' $U_t(1)$ rotation respectively
\beq\label{u-generators}
\begin{array}{c}
\wt{\mr T}_\alpha\vp{\wt{\mr T}}^\beta=\bar Z_\alpha Z^\beta-\frac14\de^\beta_\alpha(\bar ZZ),\quad\wt{\mr T}_i\vp{\wt{\mr T}}^j=\bar\xi_i\xi^j-\frac14\de_i^j(\bar\xi\xi),\\[0.2cm]
\mr Q_\alpha\vp{\mr Q}^j=\bar Z_\alpha\xi^j,\quad\mr Q_i\vp{\mr Q}^\beta=\bar\xi_iZ^\beta,\\[0.2cm]
\mr T=\bar ZZ+\bar\xi\xi,\quad\mr U=\bar ZZ-\bar\xi\xi.
\end{array}
\eeq
Throughout this paper tilde over a tensor indicates that it is traceless under contraction of its upper and lower indices of the same sort. Component form of $u(2,2|4)$ superalgebra relations (\ref{u224-algebra}) reads
\beq\label{comp-u224-algebra}
\begin{array}{c}
\{\wt{\mr T}_\alpha\vp{\wt{\mr T}}^\beta,\wt{\mr T}_\gamma\vp{\wt{\mr T}}^\de\}_{D.B.}=i(\de^\beta_\gamma\wt{\mr T}_\alpha\vp{\wt{\mr T}}^\de-\de_\alpha^\de\wt{\mr T}_\gamma\vp{\wt{\mr T}}^\beta),\quad
\{\wt{\mr T}_i\vp{\wt{\mr T}}^j,\wt{\mr T}_k\vp{\wt{\mr T}}^l\}_{D.B.}=i(\de^j_k\wt{\mr T}_i\vp{\wt{\mr T}}^l-\de_i^l\wt{\mr T}_k\vp{\wt{\mr T}}^j),\\[0.2cm]
\{\mr Q_\alpha\vp{\mr Q}^j,\mr Q_k\vp{\mr Q}^\de\}_{D.B.}=i(\de^j_k\wt{\mr T}_\alpha\vp{\wt{\mr T}}^\de+\de^\de_\alpha\wt{\mr T}_k\vp{\wt{\mr T}}^j+\frac14\de_\alpha^\de\de^j_k\mr T),\\[0.2cm]
\{\wt{\mr T}_\alpha\vp{\wt{\mr T}}^\beta,\mr Q_\gamma\vp{\mr Q}^l\}_{D.B.}=i(\de^\beta_\gamma\mr Q_\alpha\vp{\mr Q}^l-\frac14\de^\beta_\alpha\mr Q_\gamma\vp{\mr Q}^l),\quad
\{\wt{\mr T}_\alpha\vp{\wt{\mr T}}^\beta,\mr Q_k\vp{\mr Q}^\de\}_{D.B.}=-i(\de^\de_\alpha\mr Q_k\vp{\mr Q}^\beta-\frac14\de^\beta_\alpha\mr Q_k\vp{\mr Q}^\de),\\[0.2cm]
\{\wt{\mr T}_i\vp{\wt{\mr T}}^j,\mr Q_\gamma\vp{\mr Q}^l\}_{D.B.}=-i(\de^l_i\mr Q_\gamma\vp{\mr Q}^j-\frac14\de^j_i\mr Q_\gamma\vp{\mr Q}^l),\quad
\{\wt{\mr T}_i\vp{\wt{\mr T}}^j,\mr Q_k\vp{\mr Q}^\de\}_{D.B.}=i(\de^j_k\mr Q_i\vp{\mr Q}^\de-\frac14\de^j_i\mr Q_k\vp{\mr Q}^\de),\\[0.2cm]
\{\mr U,\mr Q_\alpha\vp{\mr Q}^j\}_{D.B.}=2i\mr Q_\alpha\vp{\mr Q}^j,\quad\{\mr U,\mr Q_i\vp{\mr Q}^\beta\}_{D.B.}=-2i\mr Q_i\vp{\mr Q}^\beta.
\end{array}
\eeq
Few remarks regarding above relations are in order. $su(2,2)\oplus su(4)$ and supersymmetry generators span $psu(2,2|4)$ -- the minimal superalgebra that includes conformal and $R-$symmetries. To obtain in closed form corresponding (anti)commutation relations it is common to set $\mr T=0$. The generators of $psu(2,2|4)$ and $\mr T$ span $su(2,2|4)$ superalgebra. Unlike the case $N\not=4$ this superalgebra is not simple since $\mr T$ forms an Abelian ideal. Because $U$ does not appear on the r.h.s. of (\ref{comp-u224-algebra}) $u(2,2|4)$ superalgebra has the structure of semidirect sum of $su(2,2|4)$ and $u_t(1)$.

Component form of $U(2,2|4)$ transformations (\ref{u224-stwistor-transformation}) is obtained by calculating D.B. of the supertwistor components with the individual generators in (\ref{u-generators}). In such a way we find transformation rules of the supertwistor components under $SU(2,2)\times SU(4)$ rotations
\beq
\begin{array}{c}
\de Z^\alpha=i\Lambda^\alpha\vp{\Lambda}_\beta Z^\beta,\quad\de\bar Z_\alpha=-i\bar Z_\beta\Lambda^\beta\vp{\Lambda}_\alpha,\quad \Lambda^\alpha\vp{\Lambda}_\alpha=0;\\[0.2cm]
\de\xi^i=i\Lambda^i\vp{\Lambda}_j\xi^j,\quad\de\bar\xi_i=-i\bar\xi_j\Lambda^j\vp{\Lambda}_i,\quad\Lambda^i\vp{\Lambda}_i=0,
\end{array}
\eeq
supersymmetry transformations
\beq
\de Z^\alpha=i\varepsilon^\alpha\vp{\varepsilon}_i\xi^i,\quad\de\bar\xi_i=-i\bar Z_\alpha\varepsilon^\alpha\vp{\varepsilon}_i;\quad
\de\bar Z_\alpha=-i\bar\xi_i\bar\varepsilon^i\vp{\bar\varepsilon}_\alpha,\quad\de\xi^i=i\bar\varepsilon^i\vp{\bar\varepsilon}_\alpha Z^\alpha,\quad
(\varepsilon^\alpha\vp{\varepsilon}_i)^\dagger=\bar\varepsilon^i\vp{\bar\varepsilon}_\alpha,
\eeq
as well as, $U(1)$ and $U_t(1)$ rotations
\beq\label{scaling}
\de Z^\alpha=iaZ^\alpha,\quad\de\bar Z_\alpha=-ia\bar Z_\alpha,\quad\de\xi^i=ia\xi^i,\quad\de\bar\xi_i=-ia\bar\xi_i;
\eeq
\beq
\de Z^\alpha=ia_tZ^\alpha,\quad\de\bar Z_\alpha=-ia_t\bar Z_\alpha,\quad\de\xi^i=-ia_t\xi^i,\quad\de\bar\xi_i=ia_t\bar\xi_i.
\eeq

\subsubsection{$OSp(8|8)$ global symmetry}

Action (\ref{sparticle}) is also invariant under the symmetries
generated by monomials composed of either supertwistors or dual
supertwistors only. Linear functions $\mr G_{(1,0)}=\bar{\mc
Z}_{A}\Lambda^{A}$ and $\mr G_{(0,1)}=\bar\Lambda_{A}\mc Z^{A}$
generate constant shifts of supertwistors 
\beq 
\de\mc Z^A=\{\mc
Z^A,\mr G_{(1,0)}\}_{D.B.}=i\Lambda^{A};\quad\de\bar{\mc
Z}_A=\{\bar{\mc Z}_A,\mr G_{(0,1)}\}_{D.B.}=-i\bar\Lambda_{A}.
\eeq 
Consider also generating functions defined by valence $(2,0)$
and $(0,2)$ supertwistors 
\beq 
\mr G_{(2,0)}=\bar{\mc
Z}_{A_1}\bar{\mc Z}_{A_2}\Lambda^{A_2A_1}=\bar{\mc
Z}_{A(2)}\Lambda^{A(2)},\quad\mr G_{(0,2)}=\bar\Lambda_{A_2A_1}\mc
Z^{A_1}\mc Z^{A_2}=\bar\Lambda_{A(2)}\mc Z^{A(2)}, 
\eeq 
where
convenient notation to be widely used below is $\mc Z^{A(l)}=\mc
Z^{A_1}\cdots\mc Z^{A_l}$ ($\mc Z^{A(0)}=1$, $\mc Z^{A(1)}=\mc
Z^A$) and $\bar{\mc Z}_{A(l)}=\bar{\mc Z}_{A_1}\cdots\bar{\mc
Z}_{A_l}$ ($\bar{\mc Z}_{A(0)}=1$, $\bar{\mc Z}_{A(1)}=\bar{\mc
Z}_A$).\footnote{Composite objects like $\mc Z^{A(l)}$ and
$\bar{\mc Z}_{A(l)}$ are graded symmetric in their indices. In
general it is assumed graded symmetry in supertwistor indices
denoted by the same letters. Similarly one defines the products of
supertwistor bosonic and fermionic components as 
$Z^{\alpha(m)}=Z^{\alpha_1}\cdots Z^{\alpha_m}$, $\bar
Z_{\alpha(m)}=\bar Z_{\alpha_1}\cdots\bar Z_{\alpha_m}$ and
$\xi^{i[n]}=\xi^{i_1}\cdots\xi^{i_n}$,
$\bar\xi_{i[n]}=\bar\xi_{i_1}\cdots\bar\xi_{i_n}$ ($n\leq N=4$) that
are (anti)symmetric. Antisymmetry in a set of $n$ indices is
indicated by placing $n$ in square brackets. Both symmetrization
and antisymmetrization are performed with unit weight.}
Associated variations of supertwistors read
\beq\label{osp-u224-transformations} 
\de\mc Z^A=\{\mc Z^A,\mr
G_{(2,0)}\}_{D.B.}=2i\bar{\mc Z}_B\Lambda^{BA};\quad\de\bar{\mc
Z}_A=\{\bar{\mc Z}_A,\mr G_{(0,2)}\}_{D.B.}=-2i\bar\Lambda_{AB}\mc
Z^B. 
\eeq 
Corresponding Noether currents can be identified with
valence $(2,0)$ and $(0,2)$ supertwistors
\beq\label{osp-u224-generators} \mr T_{A(2)}=\bar{\mc
Z}_{A_1}\bar{\mc Z}_{A_2},\quad\mr T^{A(2)}=\mc Z^{A_1}\mc
Z^{A_2}. \eeq 
Together with $u(2,2|4)$ currents
(\ref{u224-stwistor'}) they generate $OSp(8|8)$ global symmetry of
the Shirafuji model (\ref{sparticle}). (Anti)commutation relations
of $osp(8|8)$ superalgebra are given by (\ref{u224-algebra}) and
\beq\label{osp-u224-relations}
\begin{array}{rl}
\{\mr T_{A(2)},\mr T^{B(2)}\}_{D.B.}=&-i\left((-)^{a_2}\de^{B_1}_{A_2}\mr T_{A_1}\vp{\mr T}^{B_2}+(-)^{a_2(b_1+b_2)}\de^{B_2}_{A_2}\mr T_{A_1}\vp{\mr T}^{B_1}\right.\\[0.2cm]
+&\left.(-)^{b_1(a_1+a_2)}\de^{B_1}_{A_1}\mr T_{A_2}\vp{\mr T}^{B_2}+(-)^{a_2b_2+a_1(b_1+b_2)}\de^{B_2}_{A_1}\mr T_{A_2}\vp{\mr T}^{B_1}\right),\\[0.2cm]
\{\mr T_{A(2)},\mr T_B\vp{\mr T}^C\}_{D.B.}=&-i\left((-)^{a_2\varepsilon_b^c}\de^C_{A_2}\mr T_{A_1B}+(-)^{a_1\varepsilon_b^c+a_2c}\de^C_{A_1}\mr T_{A_2B}\right),\\[0.2cm]
\{\mr T^{A(2)},\mr T_B\vp{\mr T}^C\}_{D.B.}=&i\left(\de^{A_2}_B\mr T^{A_1C}+(-)^{a_2b}\de^{A_1}_B\mr T^{A_2C}\right).
\end{array}
\eeq

Decomposition of the generators (\ref{osp-u224-generators})
\beq\label{osp-u224-generators-comp}
\begin{array}{c}
\mr T_{AB}=\{\mr T_{\alpha\beta},\ \mr T_{ij};\ \mr Q_{\alpha j}\}:\ \mr T_{\alpha\beta}=\bar Z_\alpha\bar Z_\beta,\ \mr T_{ij}=\bar\xi_i\bar\xi_j,\ \mr Q_{\alpha j}=\bar Z_\alpha\bar\xi_j;\\[0.2cm]
\mr T^{AB}=\{\mr T^{\alpha\beta},\ \mr T^{ij};\ \mr Q^{\alpha j}\}:\ \mr T^{\alpha\beta}=Z^\alpha Z^\beta,\ \mr T^{ij}=\xi^i\xi^j,\ \mr Q^{\alpha j}=Z^\alpha\xi^j
\end{array}
\eeq
allows to find component form of the transformations (\ref{osp-u224-transformations})
\beq
\begin{array}{c}
\de Z^\alpha=2i\Lambda^{\alpha\beta}\bar Z_\beta;\\[0.2cm]
\de Z^\alpha=-2i\Lambda^{\alpha i}\bar\xi_i,\quad\de\xi^i=2i\Lambda^{\alpha i}\bar Z_\alpha;\\[0.2cm]
\de\xi^i=-2i\Lambda^{ij}\bar\xi_j
\end{array}
\eeq
and
\beq
\begin{array}{c}
\de\bar Z_\alpha=-2i\bar\Lambda_{\alpha\beta}Z^\beta;\\[0.2cm]
\de\bar Z_\alpha=-2i\bar\Lambda_{\alpha i}\xi^i,\quad\de\bar\xi_i=-2i\bar\Lambda_{\alpha i}Z^\alpha;\\[0.2cm]
\de\bar\xi_i=-2i\bar\Lambda_{ij}\xi^j.
\end{array}
\eeq
Component form of the relations (\ref{osp-u224-relations}) that involve $osp(8|8)\setminus u(2,2|4)$ currents reads
\beq\label{osp-osp-sparticle}
\begin{array}{c}
\{\mr T_{\alpha\beta},\mr T^{\gamma\delta}\}_{D.B.}=-i(\de^\gamma_\beta\wt{\mr T}_\alpha\vp{\wt{\mr T}}^\de+\de^\de_\beta\wt {\mr T}_\alpha\vp{\wt{\mr T}}^\gamma+\de^\gamma_\alpha\wt{\mr T}_\beta\vp{\wt{\mr T}}^\de+\de^\de_\alpha\wt{\mr T}_\beta\vp{\wt{\mr T}}^\gamma)
-\frac{i}{4}(\de^\gamma_\alpha\de^\de_\beta+\de^\de_\alpha\de^\gamma_\beta)(\mr T+\mr U),\\[0.2cm]
\{\mr T_{\alpha\beta},\mr Q^{\gamma l}\}_{D.B.}=-i(\de^\gamma_\beta\mr Q_\alpha\vp{\mr Q}^l+\de^\gamma_\alpha\mr Q_\beta\vp{\mr Q}^l),\\[0.2cm]
\{\mr T_{ij},\mr T^{kl}\}_{D.B.}=i(\de^k_j\wt{\mr T}_i\vp{\wt{\mr T}}^l-\de^l_j\wt{\mr T}_i\vp{\wt{\mr T}}^k-\de^k_i\wt{\mr T}_j\vp{\wt{\mr T}}^l+\de^l_i\wt{\mr T}_j\vp{\wt{\mr T}}^k)
+\frac{i}{4}(\de^l_i\de^k_j-\de^k_i\de^l_j)(\mr T-\mr U),\\[0.2cm]
\{\mr T_{ij},\mr Q^{\gamma l}\}_{D.B.}=i(\de^l_j\mr Q_i\vp{\mr Q}^\gamma-\de^l_i\mr Q_j\vp{\mr Q}^\gamma),\\[0.2cm]
\{\mr Q_{\alpha j},\mr T^{\gamma\delta}\}_{D.B.}=-i(\de^\gamma_\alpha\mr Q_j\vp{\mr Q}^\de+\de^\de_\alpha\mr Q_j\vp{\mr Q}^\gamma),\quad
\{\mr Q_{\alpha j},\mr T^{kl}\}_{D.B.}=i(\de^k_j\mr Q_\alpha\vp{\mr Q}^l-\de^l_j\mr Q_\alpha\vp{\mr Q}^k),\\[0.2cm]
\{\mr Q_{\alpha j},\mr Q^{\gamma l}\}_{D.B.}=i(\de^l_j\wt{\mr T}_\alpha\vp{\wt{\mr T}}^\gamma-\de^\gamma_\alpha\wt{\mr T}_j\vp{\wt{\mr T}}^l+\frac{1}{4}\de^l_j\de^\gamma_\alpha\mr U).
\end{array}
\eeq
Accordingly D.B. relations involving both $u(2,2|4)$ and $osp(8|8)\setminus u(2,2|4)$ generators acquire the form
$$
\{\mr T_{\alpha\beta},\wt{\mr T}_\gamma\vp{\wt{\mr T}}^\de\}_{D.B.}=-i\left(\de^\de_\beta\mr T_{\alpha\gamma}+\de^\de_\alpha\mr T_{\beta\gamma}-\frac12\de^\de_\gamma\mr T_{\alpha\beta}\right), 
$$ 
$$
\{\mr T_{\alpha\beta},\mr T\}_{D.B.}=\{\mr T_{\alpha\beta},\mr U\}_{D.B.}=-2i\mr T_{\alpha\beta}, 
$$ 
$$
\{\mr T_{\alpha\beta},\mr Q_k\vp{\mr Q}^\de\}_{D.B.}=-i(\de^\de_\beta\mr Q_{\alpha k}+\de^\de_\alpha\mr Q_{\beta k}), 
$$ 
$$
\{\mr T_{ij},\wt{\mr T}_k\vp{\wt{\mr T}}^l\}_{D.B.}=-i\left(\de^l_j\mr T_{ik}+\de^l_i\mr T_{kj}-\frac12\de^l_k\mr T_{ij}\right), 
$$ 
\beq\label{osp-u224-comp} 
\{\mr T_{ij},-\mr T\}_{D.B.}=\{\mr T_{ij},\mr U\}_{D.B.}=2i\mr T_{ij}, 
\eeq 
$$ 
\{\mr T_{ij},\mr Q_\gamma\vp{\mr Q}^l\}_{D.B.}=i(\de^l_j\mr Q_{\gamma i}-\de^l_i\mr Q_{\gamma j}), 
$$ 
$$
\{\mr Q_{\alpha j},\wt{\mr T}_\gamma\vp{\wt{\mr T}}^\de\}_{D.B.}=-i\left(\de^\de_\alpha\mr Q_{\gamma j}-\frac14\de^\de_\gamma\mr Q_{\alpha j}\right),\quad
\{\mr Q_{\alpha j},\mr T\}_{D.B.}=-2i\mr Q_{\alpha j}, 
$$ 
$$
\{\mr Q_{\alpha j},\mr Q_\gamma\vp{\mr Q}^l\}_{D.B.}=i\de^l_j\mr T_{\alpha\gamma},\quad
\{\mr Q_{\alpha j},\mr Q_k\vp{\mr Q}^\de\}_{D.B.}=-i\de^\de_\alpha\mr T_{jk}, 
$$ 
$$
\{\mr Q_{\alpha j},\wt{\mr T}_k\vp{\wt{\mr T}}^l\}_{D.B.}=-i\left(\de^l_j\mr Q_{\alpha k}-\frac14\de^l_k\mr Q_{\alpha j}\right),
$$
and
\beq\label{osp-u224-comp2}
\begin{array}{c}
\{\mr T^{\alpha\beta},\wt{\mr T}_\gamma\vp{\wt{\mr T}}^\de\}_{D.B.}=i\left(\de^\beta_\gamma\mr T^{\alpha\de}+\de^\alpha_\gamma\mr T^{\beta\de}-\frac12\de^\de_\gamma\mr T^{\alpha\beta}\right),\\[0.2cm]
\{\mr T^{\alpha\beta},\mr T\}_{D.B.}=\{\mr T^{\alpha\beta},\mr U\}_{D.B.}=2i\mr T^{\alpha\beta},\\[0.2cm]
\{\mr T^{\alpha\beta},\mr Q_\gamma\vp{\mr Q}^l\}_{D.B.}=i(\de^\beta_\gamma\mr Q^{\alpha l}+\de^\alpha_\gamma\mr Q^{\beta l}),\\[0.2cm]
\{\mr T^{ij},\wt{\mr T}_k\vp{\wt{\mr T}}^l\}_{D.B.}=i\left(\de^j_k\mr T^{il}+\de^i_k\mr T^{lj}-\frac12\de^l_k\mr T^{ij}\right),\\[0.2cm]
\{\mr T^{ij},\mr T\}_{D.B.}=\{\mr T^{ij},-\mr U\}_{D.B.}=2i\mr T^{ij},\\[0.2cm]
\{\mr T^{ij},\mr Q_k\vp{\mr Q}^\de\}_{D.B.}=i(\de^j_k\mr Q^{\de i}-\de^i_k\mr Q^{\de j}),\\[0.2cm]
\{\mr Q^{\alpha j},\wt{\mr T}_\gamma\vp{\wt{\mr T}}^\de\}_{D.B.}=i\left(\de^\alpha_\gamma\mr Q^{\de j}-\frac14\de^\de_\gamma\mr Q^{\alpha j}\right),\quad
\{\mr Q^{\alpha j},\mr T\}_{D.B.}=2i\mr Q^{\alpha j},\\[0.2cm]
\{\mr Q^{\alpha j},\mr Q_\gamma\vp{\mr Q}^l\}_{D.B.}=i\de^\alpha_\gamma\mr T^{jl},\quad
\{\mr Q^{\alpha j},\mr Q_k\vp{\mr Q}^\de\}_{D.B.}=i\de^j_k\mr T^{\alpha\de},\\[0.2cm]
\{\mr Q^{\alpha j},\wt{\mr T}_k\vp{\wt{\mr T}}^l\}_{D.B.}=i\left(\de^j_k\mr Q^{\alpha l}-\frac14\de^l_k\mr Q^{\alpha j}\right).
\end{array}
\eeq
One observes that the $u_t(1)$ generator $U$ appears on the r.h.s. of (\ref{osp-osp-sparticle}) in addition to all the $su(2,2|4)$ generators.
As a digression let us note that $OSp(2N|8)\supset U(2,2|N)$ symmetry is manifest in the superparticle models \cite{BL}, \cite{BLS} formulated in generalized superspace with the bosonic coordinates described by symmetric $4\times 4$ matrix that in addition to Minkowski 4-coordinates includes 6 coordinates described by the second rank antisymmetric tensor. Lagrangians of these models can be naturally written in terms of orthosymplectic supertwistors transforming linearly under $OSp(2N|8)$.\footnote{String models formulated in terms of $OSp(N|8)$ supertwistors were considered in \cite{ZU}-\cite{BdAPV}. In Ref.~\cite{BdAM} was discussed their relation to the Berkovits twistor string model \cite{Berkovits}.}

\subsubsection{Higher-spin symmetries}

The complete global symmetry of the model (\ref{sparticle}) is infinite-dimensional. Generic form of the generating function for both finite-dimensional and higher-spin symmetries is
\beq
\mr G_{(k,l)}=\bar{\mc Z}_{B_1}\cdots\bar{\mc Z}_{B_k}\Lambda^{B_k\cdots B_1}\vp{\Lambda}_{A_l\cdots A_1}\mc Z^{A_1}\cdots\mc Z^{A_l},\quad k,l\geq0.
\eeq
Parameters $\Lambda^{B(k)}\vp{\Lambda}_{A(l)}$ (anti)commute with themselves and with the supertwistor components depending on their parities defined by the sum $\varepsilon(\Lambda)=\sum\limits_kb_k+\sum\limits_la_l$ of parities $b_k$ and $a_l$ that take values 0(1) for the indices corresponding to bosonic (fermionic) components of supertwistors. Using associated variation of supertwistors
\beq
\begin{array}{c}
\de\mc Z^A=\{\mc Z^A,\mr G_{(k,l)}\}_{D.B.}=ik\bar{\mc Z}_{B_2}\cdots\bar{\mc Z}_{B_k}\Lambda^{B_k\cdots B_2A}\vp{\Lambda}_{C(l)}\mc Z^{C(l)},\\[0.2cm]
\de\bar{\mc Z}_A=\{\bar{\mc Z}_A,\mr G_{(k,l)}\}_{D.B.}=-il\bar{\mc Z}_{C(k)}\Lambda^{C(k)}\vp{\Lambda}_{AB_{l-1}\cdots B_1}\mc Z^{B_1}\cdots\mc Z^{B_{l-1}}
\end{array}
\eeq
one derives the variation of the superparticle's Lagrangian
\beq
\de\mathscr L=(k+l-2)\frac{i}{2}\frac{d}{d\tau}\left(\bar{\mc Z}_{A(k)}\Lambda^{A(k)}\vp{Lambda}_{B(l)}\mc Z^{B(l)}\right).
\eeq
From this expression it becomes clear that $OSp(8|8)$ symmetry for which $(k,l)=(1,1)$, $(2,0)$ or $(0,2)$ is special since the action is invariant under corresponding variation. For other values of $(k,l)$ the invariance is only up to a total divergence. So the complete infinite-dimensional symmetry of the superparticle model (\ref{sparticle}) is generated by the sum
\beq\label{generatingfunction}
\mr G=\sum\limits_{k,l\geq0}\mr G_{(k,l)}=\sum\limits_{k,l\geq0}\bar{\mc Z}_{B(k)}\Lambda^{B(k)}\vp{\Lambda}_{A(l)}\mc Z^{A(l)}.
\eeq
Associated Noether currents are given by a collection of all possible monomials of the form
\beq\label{monomials}
\mr T^{(k,l)}\vp{\mr T}_{A(k)}\vp{\mr T}^{B(l)}=\bar{\mc Z}_{A(k)}\mc Z^{B(l)},\quad k,l\geq0.
\eeq
Such monomials span an infinite-dimensional superalgebra, whose (anti)commutation relations in schematic form read
\beq\label{acl-relations}
\begin{array}{rl}
\{\mr T^{(k,l)}\vp{\mr T}_{B(k)}\vp{\mr T}^{A(l)},\mr T^{(p,q)}\vp{\mr T}_{D(p)}\vp{\mr T}^{C(q)}\}_{D.B.}=&i(\de^{A(1)}_{D(1)}\mr T^{(k+p-1,l+q-1)}\vp{\mr T}_{B(k)D(p-1)}\vp{\mr T}^{A(l-1)C(q)}\\[0.2cm]
-&\de^{C(1)}_{B(1)}\mr T^{(k+p-1,l+q-1)}\vp{\mr T}_{B(k-1)D(p)}\vp{\mr T}^{A(l)C(q-1)}).
\end{array}
\eeq

\subsection{Quantum symmetries of $D=4$ $N=4$ massless superparticle}

At the quantum level D.B. relations (\ref{DirBr}) are replaced by
(anti)commutators\footnote{Planck constant is omitted on the
r.h.s.} 
\beq\label{quantizedsupertwistors} 
[\hat{\mc
Z}^A,\hat{\bar{\mc Z}}_B\}=\de^A_B,\quad [\hat{\bar{\mc
Z}}_B,\hat{\mc Z}^A\}=-(-)^a\de^A_B 
\eeq 
and components of
supertwistors and their duals become Hermitian conjugate operators
$(\hat{\mc Z}^A)^\dagger=\hat{\bar{\mc Z}}_A$.\footnote{Hermitian
conjugation is assumed to reverse the order for both bosonic and
fermionic operators.} Thus global symmetry generators are promoted
to Hermitian operators. Quantized $u(2,2|4)$ generators
$\widehat{\mr T}_A\vp{\hat{\mr T}}^B$ are defined by the graded
symmetrized (Weyl ordered) expression 
\beq 
\widehat{\mr
T}_A\vp{\hat{\mr T}}^B=\frac12(\hat{\bar{\mc Z}}_A\hat{\mc
Z}^B+(-)^{ab}\hat{\mc Z}^B\hat{\bar{\mc Z}}_A). 
\eeq 
As far as
component generators (\ref{u224-generators-comp}),
(\ref{u-generators}) are concerned there are no ambiguities in the
definition of $su(2,2)$ and $su(4)$ generators, because of their
tracelessness, and supersymmetry generators, while $u(1)$ and
$u_t(1)$ generators can be presented in various forms
\beq\label{t-quantum} 
\widehat{\mr T}=\frac12(\hat{\bar Z}\hat
Z+\hat Z\hat{\bar
Z})+\frac12(\hat{\bar\xi}\hat\xi-\hat\xi\hat{\bar\xi})=\hat{\bar
Z}\hat Z+\hat{\bar\xi}\hat\xi=\hat Z\hat{\bar
Z}-\hat\xi\hat{\bar\xi}, 
\eeq 
\beq 
\widehat{\mr
U}=\frac12(\hat{\bar Z}\hat Z+\hat Z\hat{\bar
Z})-\frac12(\hat{\bar\xi}\hat\xi-\hat\xi\hat{\bar\xi})=\hat{\bar
Z}\hat Z-\hat{\bar\xi}\hat\xi+4=\hat Z\hat{\bar
Z}+\hat\xi\hat{\bar\xi}-4. 
\eeq 
There are no numerical constants
in 'asymmetric' representations of $\widehat{\mr T}$ since equal
in number bosonic and fermionic components of supertwistor and its
dual give contributions that cancel each other. Also no ambiguity
arises in the definition of quantized $osp(8|8)\setminus u(2,2|4)$
generators (\ref{osp-u224-generators}). In general higher-spin
generators $\widehat{\mr T}^{(k,l)}\vp{\mr T}_{A(k)}\vp{\mr
T}^{B(l)}$ are defined by the sum of all graded permutations of
constituent supertwistors 
\beq\label{q-generators} 
\widehat{\mr
T}^{(k,l)}\vp{\mr T}_{B(k)}\vp{\mr
T}^{A(l)}=\frac{1}{(k+l)!}\left(\hat{\bar{\mc Z}}_{B(k)}\hat{\mc
Z}^{A(l)}+\cdots\right). 
\eeq 
They satisfy (anti)commutation
relations \cite{FrLin-AP90} that in schematic form read
\beq\label{q-relations}
\begin{array}{c}
[\widehat{\mr T}^{(k,l)}\vp{\widehat{\mr T}}_{B(k)}\vp{\mr T}^{A(l)},\widehat{\mr T}^{(p,q)}\vp{\widehat{\mr T}}_{D(p)}\vp{\mr T}^{C(q)}\} \\[0.2cm]
\sim\sum\limits_{\substack{m,n\geq0 \\ m+n\ \mathrm{odd}}}^{\substack{m\leq\mathrm{min}(l,p)\\ n\leq\mathrm{min}(k,q)}}\de^{A(m)}_{D(m)}\de^{C(n)}_{B(n)}\widehat{\mr T}^{(k+p-m-n,l+q-m-n)}\vp{\widehat{\mr T}}_{B(k-n)D(p-m)}\vp{\widehat{\mr T}}^{A(l-m)C(q-n)}.
\end{array}
\eeq

\subsection{Higher-spin symmetries of massless superparticle and higher-spin superalgebras}

It is worthwhile to compare considered classical and quantum relations
of the higher-spin currents with those of the higher-spin superalgebras
based on orthosymplectic symmetries \cite{Vasiliev-FP88}.
Generating function (\ref{generatingfunction}) of the
infinite-dimensional global symmetry of the superparticle action
(\ref{sparticle}) can be considered as a symbol of the operator
$\widehat{\mr G}$ that is defined by the same expression
(\ref{generatingfunction}) in which (Weyl ordered products of)
quantized supertwistors (\ref{quantizedsupertwistors}) should be
substituted. Associative algebra $aq(8|8)$ (aq='associative
quantum') of such operators is isomorphic to the $\ast$-product
algebra of their symbols. The $\ast$-product can be brought to the
following form in terms of $PSU(2,2|4)$ supertwistors
\cite{FrLin-AP90} 
\beq A(\mc Z,\bar{\mc Z})\ast B(\mc Z,\bar{\mc
Z})=Ae^{\Delta}B,\quad\Delta=\left(\frac{\overleftarrow{\partial}}{\partial\mc
Z}\frac{\overrightarrow{\partial}}{\partial\bar{\mc
Z}}-\frac{\overrightarrow{\partial}}{\partial\mc
Z}\frac{\overleftarrow{\partial}}{\partial\bar{\mc Z}}\right),
\eeq 
where $A(\mc Z,\bar{\mc Z})$ and $B(\mc Z,\bar{\mc Z})$ are
symbols of the operators $\widehat A(\hat{\mc Z},\hat{\bar{\mc
Z}})$ and $\widehat B(\hat{\mc Z},\hat{\bar{\mc Z}})$.
Introduction of the Lie superalgebra structure in $aq(8|8)$
requires assignment of parities to the monomials $\widehat{\mr
T}^{(k,l)}\vp{\widehat{\mr T}}_{B(k)}\vp{\mr T}^{A(l)}$ (and associated
expansion coefficients) in $\mr G$. The prescription
\cite{Vasiliev-FP88} appropriate for the construction of
higher-spin gauge theories consists in ascribing parity 1(0) to
$SU(2,2)$ ($SU(4)$) indices. Such a choice agrees with the
spin-statistics relation for the expansion coefficients that are
identified with the potentials (field strengths) of higher-spin
gauge fields but, for instance, the generators of such a Lie
superalgebra defined by the product of an odd number of
supertwistor bosonic components should satisfy anticommutation relations,
while those equal the product of supertwistor fermionic components
-- commutation relations. We adhere to alternative prescription
motivated by the symmetries of the superparticle action that,
however, results in wrong spin-statistics relation for some of the
parameters in (\ref{generatingfunction}). Both prescriptions match
for the superalgebras spanned by the generators composed of an
even number of supertwistors, in particular for the
finite-dimensional symmetries generated by quadratic monomials.

The r.h.s. of (anti)commutators $A\ast
B-(-)^{\varepsilon(A)\varepsilon(B)}B\ast A$ of $aq(8|8)$ elements 
for the chosen parity assignment prescription coincides with 
that of quantized generators
(\ref{q-generators}) 
so that relations (\ref{q-relations}) can be identified as
corresponding to infinite-dimensional Lie superalgebra
$Lie[aq](8|8)$. Then D.B. relations (\ref{acl-relations}) can be
identified with the classical limit $\hbar\to0$ of
$Lie[aq](8|8)$ that can be named $Lie[acl](8|8)$. In general the
classical limit of Lie superalgebras based on $\ast-$product
associative algebras is obtained by introducing explicit
dependence on $\hbar$ \beq \left(A\ast
B-(-)^{\varepsilon(A)\varepsilon(B)}B\ast A\right)(\mc Z,\bar{\mc
Z})=\frac{1}{\hbar}\left(Ae^{\Delta_\hbar}B-(-)^{\varepsilon(A)\varepsilon(B)}Be^{\Delta_\hbar}A\right),\quad\Delta_\hbar=\hbar\Delta
\eeq and taking $\hbar\to0$. $\hbar$ plays the role of the
contraction parameter as was explained in \cite{FrVasAP87},
\cite{Vasiliev-FP88}.


\subsection{$D=4$ $N=4$ massless superparticle with gauged $U(1)$ symmetry}

Modification of the model (\ref{sparticle}) considered by Shirafuji \cite{Shirafuji} consists in gauging $U(1)$ symmetry (\ref{scaling}) by adding $\mr T$ with the Lagrange multiplier to the action
\beq\label{sparticle2}
S_{U(1)}=\int d\tau\lambda\mr T.
\eeq
Gauging phase symmetry means that superparticle propagates on projective supertwistor space. Global symmetry of the action (\ref{sparticle}), (\ref{sparticle2}) is described by a subalgebra of $Lie[acl](8|8)$ spanned by the generators commuting with $\mr T$. We identify this superalgebra as $iu_{cl}(2,2|4)$. It is the classical limit of $iu(2,2|4)$ (iu='infinite-dimensional unitary') superalgebra introduced in \cite{FrLin-AP90}.
Both $iu(2,2|4)$ and $iu_{cl}(2,2|4)$ share the same finite-dimensional subalgebra $u(2,2|4)$.
$iu_{cl}(2,2|4)$ is the centralizer of $\mr T$ in $Lie[acl](8|8)$ analogously to the case of corresponding Lie superalgebras based on associative $\ast-$product algebras (see, e.g., relevant discussion in \cite{Vasiliev-PRD01}). Generators of $iu_{cl}(2,2|4)$ are the same as those of $iu(2,2|4)$ and are given by the monomials in (\ref{monomials}) with equal number of supertwistors and dual supertwistors
\beq\label{monomials2}
\mr T^{(L,L)}\vp{\mr T}_{A(L)}\vp{\mr T}^{B(L)}\equiv\mr T^{(L)}\vp{\mr T}_{A(L)}\vp{\mr T}^{B(L)}=\bar{\mc Z}_{A(L)}\mc Z^{B(L)},\quad L\geq1.
\eeq
Number $L$ we shall call the level of the generator following \cite{FrLin-NPB91}. D.B. relations of the generators (\ref{monomials2}) can be derived from (\ref{acl-relations})
\beq
\begin{array}{rl}
\{\mr T^{(L_1)}\vp{\mr T}_{B(L_1)}\vp{\mr T}^{A(L_1)},\mr T^{(L_2)}\vp{\mr T}_{D(L_2)}\vp{\mr T}^{C(L_2)}\}_{D.B.}=&i(\de^{A(1)}_{D(1)}\mr T^{(L_1+L_2-1)}\vp{\mr T}_{B(L_1)D(L_2-1)}\vp{\mr T}^{A(L_1-1)C(L_2)}\\[0.2cm]
-&\de^{C(1)}_{B(1)}\mr T^{(L_1+L_2-1)}\vp{\mr T}_{B(L_1-1)D(L_2)}\vp{\mr T}^{A(L_1)C(L_2-1)}).
\end{array}
\eeq

Going on the constraint shell $\mr T\approx0$ implies setting to zero those generators of $iu_{cl}(2,2|4)$ that are multiples of $\mr T$. This narrows down higher-spin symmetry to the subalgebra of $iu_{cl}(2,2|4)$. For $N\not=4$ such a symmetry is generated by the classical limit of $isl(2,2|N)$ superalgebra \cite{FrLin-AP90}. The construction of this superalgebra is based on the direct sum representation of $u(2,2|N)$ as $su(2,2|N)\oplus\mr T_N$ \footnote{In this and some of the subsequent formulas to avoid confusion $\mr T$ is endowed with the subscript explicitly indicating the number of odd components of associated supertwistors.} and its higher-spin generalization. The $N=4$ case requires special treatment since $su(2,2|4)=psu(2,2|4)\oplus\mr T_4$.

\subsection{Spectrum identification}

Important realization \cite{Penrose68}, \cite{Shirafuji} of quantized (super)twistors related to the definition of twistor wave functions is to treat components of $\hat{\mc Z}^A$ as classical quantities, while components of the dual supertwistor $\hat{\bar{\mc Z}}_A$ are considered as differential operators
\beq\label{derivatives}
\hat{\bar Z}_\alpha\to-\frac{\partial}{\partial Z^\alpha},\quad\hat{\bar\xi}_i\to\frac{\vec\partial}{\partial\xi^i}
\eeq
acting in the space of (homogeneous) functions of $(Z,\xi)$. Alternatively, components of $\hat{\bar{\mc Z}}_A$ may be treated as $c-$numbers, while components of $\hat{\mc Z}^A$ are replaced by differential operators.

Quantum generator (\ref{t-quantum}) of $U(1)$ phase symmetry in the realization (\ref{derivatives}) can be brought to the form
\beq\label{qT}
\widehat{\mr T}=2\hat{\mr s}+2-\xi\frac{\partial}{\partial\xi},
\eeq
where $\hat{\mr s}=-1-\frac12Z\frac{\partial}{\partial Z}$ is the helicity operator \cite{PenR}. Any homogeneous function on the supertwistor space is its eigenfunction including the wave function $F(Z,\xi)$ of the superparticle propagating on the projective supertwistor space that satisfies
\beq\label{helicity-eq}
\left(\hat{\mr s}+1-\frac12\xi\frac{\partial}{\partial\xi}\right)F(Z,\xi)=0.
\eeq
The solution to this equation is
\beq
F(Z,\xi)=f(Z)+\xi^i\varphi_{i}(Z)+\frac{1}{2!}\xi^{i[2]}f_{i[2]}(Z)+\frac{1}{3!}\xi^{i[3]}\varphi_{i[3]}(Z)+\frac{1}{4!}\xi^{i[4]}f_{i[4]}(Z).
\eeq
Even components in the expansion $f$, $f_{i[2]}$ and $f_{i[4]}$ upon the twistor transform \cite{PenR} describe bosonic particles of helicities $-1$, $0$ and $+1$ on (complexified conformally-compactified) Minkowski space-time, whereas odd components $\varphi_{i}$ and $\varphi_{i[3]}$ -- fermions of helicities $-1/2$ and $+1/2$, altogether forming $D=4$ $N=4$ superYang-Mills multiplet that is CPT self-conjugate, i.e. includes particles of opposite helicities.

The spectrum of the superparticle model (\ref{sparticle}) is described by an infinite series of even $F_{2k}(Z,\xi)$ and odd $\Phi_{2k+1}(Z,\xi)$ ($k\in\mathbb Z$) eigenfunctions of the operator (\ref{qT})
\beq\label{spectrum-gen-eq}
\hspace*{-0.2cm}\left(\hat{\mr s}+1-\frac12\xi\frac{\partial}{\partial\xi}+\frac{a}{2}\right)\!\mathscr F_{a}(Z,\xi)\!=\!0\!:\!\left\{
\begin{array}{c}
\left(\hat{\mr s}+1-\frac12\xi\frac{\partial}{\partial\xi}+k\right)F_{2k}\!=\!0,\ a\!=\!2k\\
\!\!\left(\hat{\mr s}+1-\frac12\xi\frac{\partial}{\partial\xi}+k+\frac12\right)\!\Phi_{2k+1}\!=\!0,\ a\!=\!2k\!+\!1
\end{array}
\right.,
\eeq
where the subscript indicates homogeneity degree.
Pairs of functions $\mathscr F_{a}$ and $\mathscr F_{-a}$ ($a>0$) describe (upon the twistor transform) CPT conjugate doubleton supermultiplets \cite{Marcus}, \cite{GMZ} with helicities ranging from $-a/2-1$ to $a/2+1$ that are even\footnote{The supermultiplet is even/odd if the particle with the highest value of the modulus of helicity is boson/fermion.} for even values of $a$ and odd for odd values of $a$. This explains the notation introduced in Eq.~(\ref{spectrum-gen-eq}). The low-spin supermultiplets in this series are $\Phi_{\pm1}$ with particles of helicities $0$, $\pm1/2$, $\pm1$, $\pm3/2$ and $F_{\pm2}$ describing $D=4$ $N=4$ Einstein supergravity. The only exception is $F_{0}\equiv F$ twistor function corresponding to the self-conjugate superYang-Mills multiplet considered in the previous paragraph.

Another way to derive the spectrum \cite{Claus} of the
superparticle model (\ref{sparticle}) is to transform \cite{GMZ2},
\cite{Chiodaroli} supertwistor components into two pairs of
bosonic and two pairs of fermionic $SU(2)$ oscillators realizing
$[SU(2)]^4$ -- the maximal compact subgroup of $SU(2,2|4)$. They
provide minimal (single generation) oscillator realization of the
$su(2,2|4)$ superalgebra and are used to construct doubleton
supermultiplets \cite{Bars}, \cite{Marcus}, \cite{GMZ}. All these
supermultiplets assemble into two singleton supermultiplets of
$osp(8|8)$ that arise upon quantization of the massless
superparticle on superspace with bosonic $4\times4$ matrix
coordinates \cite{BLS}.

\setcounter{equation}{0}
\section{Higher-spin supersymmetries in twistor string models}

For Lorentzian signature world sheet the simplest twistor string action can be presented as
\beq\label{genBerk}
\begin{array}{c}
S=\int d\tau d\sigma(\mathscr L_L+\mathscr L_R):\\[0.2cm]
\mathscr L_L=-2(Y_\alpha\partial_-Z^\alpha+\eta_i\partial_-\xi^i)+\mathscr L_{\mathrm{L-mat}},\ \mathscr L_R=-2(\bar Y_\alpha\partial_+\bar Z^\alpha+\bar\eta_i\partial_+\bar\xi^i)+\mathscr L_{\mathrm{R-mat}},
\end{array}
\eeq
where $\partial_{\pm}=\frac12(\partial_\tau\pm\partial_\s)$, $\sigma^{\pm}=\tau\pm\s$, $Y_{+\alpha}\equiv Y_{\alpha}$, $\bar Y_{-\alpha}\equiv\bar Y_\alpha$, $\eta_{+i}\equiv\eta_i$, $\bar\eta_{-i}\equiv\bar\eta_i$ and $\mathscr L_{\mathrm{L(R)-mat}}$ are Lagrangians for left- and right-moving non-twistor matter variables, whose contribution to the world-sheet conformal anomaly equals $c=\bar c=26$ to cancel that of $(b, c)-$ghosts. Such variables may contain a current algebra for some Lie group (see, e.g., \cite{BerkWitt}).
In Berkovits twistor-string model \cite{Berkovits} global scale symmetry for both left- and right-movers
\beq\label{gl1-lr}
\begin{array}{c}
\de Z^\alpha=\Lambda Z^\alpha,\quad\de Y_\alpha=-\Lambda Y_\alpha,\quad\de\xi^i=\Lambda\xi^i,\quad\de\eta_i=-\Lambda\eta_i;\\[0.2cm]
\de\bar Z^\alpha=\bar\Lambda\bar Z^\alpha,\quad\de\bar Y_\alpha=-\bar\Lambda\bar Y_\alpha,\quad\de\bar\xi^i=\bar\Lambda\bar\xi^i,\quad\de\bar\eta_i=-\bar\Lambda\bar\eta_i
\end{array}
\eeq
is gauged by adding to the action (\ref{genBerk}) appropriate constraints $T=Y_\alpha Z^\alpha+\eta_i\xi^i\approx0$ and $\bar T=\bar Y_\alpha\bar Z^\alpha+\bar\eta_i\bar\xi^i\approx0$ with the Lagrange multipliers
\beq\label{Berk}
S_{GL(1,\mathbb R)}=\int d\tau d\sigma(\lambda T+\bar\lambda\bar T).
\eeq
This necessitates add two units to the central charges of the matter variables to compensate that of $(b, c)-$ghosts and ghosts for the gauged $GL(1,\mathbb R)$ symmetry.

Definition of the open string sector, that to date is the only one well-understood, is based on the conditions $\mc Z^A=\bar{\mc Z}^A$, $\mc Y_B=\bar{\mc Y}_B$ imposed on the world-sheet boundary on the supertwistors $\mathcal Z^A=(Z^\alpha, \xi^i)$, $\bar{\mc Z}^A=(\bar Z^\alpha, \bar\xi^i)$ and their duals $\mathcal Y_B=(Y_\beta, \eta_j)$, $\bar{\mc Y}_B=(\bar Y_\beta, \bar\eta_j)$. So taking into account reality condition of the Lagrangian one is led to consider left(right)-moving supertwistors $\mc Z^A$ ($\bar{\mc Z}^A$) and dual supertwistors $\mc Y_B$ ($\bar{\mc Y}_B$) as independent variables with real components. Such supertwistors are adapted for the description of fields on $D=4$ $N=4$ superspace for the space-time of signature $(++--).$\footnote{Detailed discussion of the reality conditions of the twistor string Lagrangian for both Lorentzian and Euclidean world sheets, and different real structures in the complex supertwistor space associated with $D=4$ space-times of various signatures can be found, e.g., in \cite{Abou-Zeid}.} Conformal group of Minkowski space-time of such a signature is $SO(3,3)\sim SL(4,\mathbb R)$ and its minimal $N=4$ supersymmetric extension is $PSL(4|4,\mathbb R)$ with the bosonic subgroup $SL(4,\mathbb R)\times SL(4,\mathbb R)$ implying that bosonic and odd components of $\mc Z^A$ belong to the fundamental representation of $SL(4,\mathbb R)_L\times SL(4,\mathbb R)_L$, whereas bosonic and odd components of $\mc Y_A$  belong to the antifundamental representation. Correspondingly components of $\bar{\mc Z}^A$ and $\bar{\mc Y}_A$ transform according to the (anti)fundamental representation of $SL(4,\mathbb R)_R\times SL(4,\mathbb R)_R$.\footnote{In this section the same letters are utilized to label supertwistors, their components and indices as in the previous one, although here they are strictly speaking different mathematical objects related to another real structure in the complex supertwistor space. For the dual supertwistor and its components we use independent notation since quantities with bars are reserved to label variables of the right-moving sector of the twistor string. We hope this will not cause a confusion since in each section only one kind of supertwistors is considered.}

Focusing on the sector of left-movers of the model (\ref{genBerk}) and applying the Dirac approach 
yields equal-time D.B. relations
\beq
\{Z^\alpha(\s),Y_\beta(\s')\}_{D.B.}=\de^\alpha_\beta\de(\sigma-\sigma'),\quad\{\xi^i(\s),\eta_j(\s')\}_{D.B.}=\de^i_j\de(\sigma-\sigma')
\eeq
that in terms of the $PSL(4|4,\mathbb R)$ supertwistors can be written as
\beq
\{\mathcal Z^A(\s),\mathcal Y_B(\s')\}_{D.B.}=\delta^A_B\de(\s-\s'),\quad\{\mathcal Y_B(\s),\mathcal Z^A(\s')\}_{D.B.}=-(-)^a\delta^A_B\de(\s-\s').
\eeq
Similar relations hold for the right-movers.

\subsection{Classical symmetries of twistor strings}

Global symmetry of the left-moving part of the action (\ref{genBerk}) is generated on D.B. by the function
\beq\label{gfglsymgenBerk1}
G=\!\int\! d\s\!\sum\limits_{L\geq0}G_{(L)}(\s),\quad G_{(L)}(\s)=\mc Y_{B}(\s)\Lambda^{B}\vp{\Lambda}_{A_L\ldots A_1}\mc Z^{A_1}(\s)\cdots\mc Z^{A_L}(\s).
\eeq 
For arbitrary value of the order $L$ transformation rules for the supertwistors read
\beq\label{gfglsymgenBerk2}
\begin{array}{c}
\delta\mc Z^A(\s)=\Lambda^{A}\vp{\Lambda}_{B(L)}\mc Z^{B(L)}(\s),\\[0.2cm]
\delta\mc Y_A(\s)=-L\mc Y_{C}(\s)\Lambda^{C}\vp{\Lambda}_{AB_{L-1}\ldots B_1}\mc Z^{B_1}(\s)\cdots\mc Z^{B_{L-1}}(\s).
\end{array}
\eeq 
Associated Noether current densities up to irrelevant
numerical factor are given by the monomials  
\beq\label{twstrNc}
T^{(L)}\vp{T}_{B}\vp{T}^{A(L)}(\s)=\mc Y_{B}\mc Z^{A(L)},\quad L\geq0
\eeq 
that enter generating functions $G_{(L)}$. On D.B. they generate the TSA\footnote{To be more precise one has to introduce TSA as an infinite-dimensional Lie superalgebra  and then consider its loop version pertinent to twistor-string global symmetry. Let us also note that the subscript L in the notation of symmetry groups and algebras will be omitted as the discussion is concentrated on the sector of left-movers only. On the boundary left- and right-moving variables are identified and thus also no subscripts are needed.} 
\beq\label{aq-relations}
\begin{array}{rl}
\{T^{(L)}\vp{T}_{B}\vp{T}^{A(L)}(\s),T^{(M)}\vp{T}_{D}\vp{T}^{C(M)}(\s')\}_{D.B.}=&(\de^{A(1)}_DT^{(L+M-1)}\vp{T}_{B}\vp{T}^{A(L-1)C(M)}
\\[0.2cm]
-&\de^{C(1)}_BT^{(L+M-1)}\vp{T}_{D}\vp{T}^{A(L)C(M-1)})(\s)\de(\s-\s').
\end{array}
\eeq

The
finite-dimensional subalgebra of $TSA$ is spanned, apart from the order 0 generator $\mc Y_A(\s)$ that is responsible for constant shift of the supertwistor components, by quadratic monomial 
\beq\label{gl-string}
T_A\vp{T}^B(\s)=\mc Y_A\mc Z^B,
\eeq
generating $gl(4|4,\mathbb R)$ superalgebra  
\beq\label{osp-rel-string}
\{T_A\vp{T}^B(\s),T_C\vp{T}^D(\s')\}_{D.B.}=(\de^B_CT_A\vp{T}^D-(-)^{\varepsilon^b_a\varepsilon^d_c}\de^D_AT_C\vp{T}^B)(\s)\de(\s-\s'),\ \varepsilon^b_a=(-)^{a+b}.
\eeq

Irreducible components of $gl(4|4,\mathbb R)$ current densities (\ref{gl-string}) are
\beq\label{gl-string-comp}
\begin{array}{c}
T_A\vp{T}^B(\s)=\{\wt T_\alpha\vp{T}^\beta,\ \wt T_i\vp{T}^j;\ Q_\alpha\vp{Q}^j,\ Q_i\vp{Q}^\beta;\ T,\ U\}:\\[0.2cm]
\wt T_\alpha\vp{T}^\beta=Y_\alpha Z^\beta-\frac14\delta_\alpha^\beta(YZ),\quad\wt T_i\vp{T}^j=\eta_i\xi^j-\frac14\de_i^j(\eta\xi);\\[0.2cm]
Q_\alpha\vp{Q}^j=Y_\alpha\xi^j,\quad Q_i\vp{Q}^\beta=\eta_i Z^\beta;\quad T=Y_\alpha Z^\alpha+\eta_i\xi^i,\quad U=Y_\alpha Z^\alpha-\eta_i\xi^i.
\end{array}
\eeq 
The densities of $sl(4,\mathbb R)\times sl(4,\mathbb R)$
currents $\wt T_\alpha\vp{T}^\beta(\s)$, $\wt T_i\vp{T}^j(\s)$ and
those of the supersymmetry currents $Q_\alpha\vp{Q}^j(\s)$,
$Q_i\vp{Q}^\beta(\s)$ span $psl(4|4,\mathbb R)$ superalgebra,
while these generators and $T(\s)$ span $sl(4|4,\mathbb R)$. On
D.B. they generate infinitesimal $SL(4,\mathbb R)\times
SL(4,\mathbb R)$ rotations of the supertwistor components
\beq
\begin{array}{c}
\de Z^\alpha(\s)=\Lambda^\alpha\vp{\Lambda}_\beta Z^\beta(\s),\quad\de Y_\alpha(\s)=-Y_\beta(\s)\Lambda^\beta\vp{\Lambda}_\alpha,\quad \Lambda^\alpha\vp{\Lambda}_\alpha=0;\\[0.2cm]
\de\xi^i(\s)=\Lambda^i\vp{\Lambda}_j\xi^j(\s),\quad\de\eta_i(\s)=-\eta_j(\s)\Lambda^j\vp{\Lambda}_i,\quad\Lambda^i\vp{\Lambda}_i=0
\end{array}
\eeq
and supersymmetry transformations
\beq
\de Z^\alpha(\s)=\varepsilon^\alpha\vp{\varepsilon}_i\xi^i(\s),\ \de\eta_i(\s)=-Y_\alpha(\s)\varepsilon^\alpha\vp{\varepsilon}_i;\quad
\de Y_\alpha(\s)=-\eta_i(\s)\epsilon^i\vp{\epsilon}_\alpha,\ \de\xi^i(\s)=\epsilon^i\vp{\epsilon}_\alpha Z^\alpha(\s),
\eeq
where $\varepsilon^\alpha\vp{\varepsilon}_i$ and $\epsilon^i\vp{\epsilon}_\alpha$ are independent odd parameters with 16 real components each.
$T(\s)$ generates $GL(1,\mathbb R)$ transformations (\ref{gl1-lr}) and $U(\s)$ -- 'twisted' $GL_t(1,\mathbb R)$ transformations
\beq
\begin{array}{c}
\de Z^\alpha(\s)=\Lambda_t Z^\alpha(\s),\ \de Y_\alpha(\s)=-\Lambda_t Y_\alpha(\s),\ \de\xi^i(\s)=-\Lambda_t\xi^i(\s),\ \de\eta_i(\s)=\Lambda_t\eta_i(\s).
\end{array}
\eeq

$gl(4|4,\mathbb R)$ relations (\ref{osp-rel-string}) are spelt out in terms of irreducible components (\ref{gl-string-comp}) as
$$
\{\wt T_\alpha\vp{T}^\beta(\s),\wt T_\gamma\vp{T}^\de(\s')\}_{D.B.}=(\de^\beta_\gamma\wt T_\alpha\vp{T}^\de-\de_\alpha^\de\wt T_\gamma\vp{T}^\beta)(\s)\de(\s-\s'),
$$
$$
\{\wt T_i\vp{\wt T}^j(\s),\wt T_k\vp{\wt T}^l(\s')\}_{D.B.}=(\de^j_k\wt T_i\vp{\wt T}^l-\de_i^l\wt T_k\vp{\wt T}^j)(\s)\de(\s-\s'),
$$
$$
\{Q_\alpha\vp{Q}^j(\s),Q_k\vp{Q}^\de(\s')\}_{D.B.}=(\de^j_k\wt T_\alpha\vp{T}^\de+\de^\de_\alpha\wt T_k\vp{T}^j+\frac14\de_\alpha^\de\de^j_k T)(\s)\de(\s-\s'),
$$
\beq\label{gl44relations}
\{\wt T_\alpha\vp{T}^\beta(\s),Q_\gamma\vp{Q}^l(\s')\}_{D.B.}=(\de^\beta_\gamma Q_\alpha\vp{Q}^l-\frac14\de^\beta_\alpha Q_\gamma\vp{Q}^l)(\s)\de(\s-\s'),
\eeq
$$
\{\wt T_\alpha\vp{T}^\beta(\s), Q_k\vp{Q}^\de(\s')\}_{D.B.}=-(\de^\de_\alpha Q_k\vp{Q}^\beta-\frac14\de^\beta_\alpha Q_k\vp{Q}^\de)(\s)\de(\s-\s'),
$$
$$
\{\wt T_i\vp{\wt T}^j(\s),Q_\gamma\vp{Q}^l(\s')\}_{D.B.}=-(\de^l_i Q_\gamma\vp{Q}^j-\frac14\de^j_iQ_\gamma\vp{Q}^l)(\s)\de(\s-\s'),
$$
$$
\{\wt T_i\vp{\wt T}^j(\s),Q_k\vp{Q}^\de(\s')\}_{D.B.}=(\de^j_kQ_i\vp{Q}^\de-\frac14\de^j_iQ_k\vp{Q}^\de)(\s)\de(\s-\s'),
$$
$$
\{U(\s),Q_\alpha\vp{Q}^j(\s')\}_{D.B.}=\!2Q_\alpha\vp{Q}^j(\s)\de(\s-\s'),\ \{U(\s),Q_i\vp{Q}^\beta(\s')\}_{D.B.}=\!-2Q_i\vp{Q}^\beta(\s)\de(\s-\s').
$$
$T(\sigma)$ commutes on D.B. with all other $gl(4|4,\mathbb R)$ current densities thus forming an Abelian ideal. The density $U(\s)$ of 'twisted' $gl_t(1,\mathbb R)$ current does not appear on the r.h.s. of (\ref{gl44relations}) that allows to consider $gl(4|4,\mathbb R)$ as the semidirect sum of $sl(4|4,\mathbb R)$ and $gl_t(1,\mathbb R)$.

\subsection{Quantum symmetries of twistor strings}

It was shown in \cite{Dolan} that $SL(4|4,\mathbb R)$ symmetry is
preserved at the quantum level, whereas the generator $U$ of
'twisted' $GL_t(1,\mathbb R)$ symmetry has anomalous OPE with the
world-sheet stress-energy tensor implying that corresponding
symmetry is broken in twistor string theory. 
Thus possible type of infinite-dimensional symmetry that
could survive in the quantum theory is restricted to that based on
$sl(4|4,\mathbb R)$ as finite-dimensional subalgebra.
Since $gl(4|4,\mathbb R)=sl(4|4,\mathbb R)\supplus gl_t(1,\mathbb R)$ superalgebra belongs to the family of $gl(M|M,\mathbb R)$ superalgebras, whose properties differ from those of $gl(M|N,\mathbb R)$ superalgebras with $M\not=N$, one is forced to take components of supertwistors as building blocks of the generators for $sl-$type superalgebras.

\subsubsection{Superalgebraic perspective on quantum higher-spin symmetries}

In the bosonic limit $TSA$ reduces to $TSA_{\mathfrak b}$ -- an infinite-dimensional Lie algebra, whose generators are obtained from (\ref{twstrNc}) by setting to zero fermionic components of the supertwistors. Order 0 and 1 generators are given by the dual bosonic twistor $Y_\alpha$ and $gl(4,\mathbb R)$ generators $Y_\alpha Z^\beta$. The latter divide into $sl(4,\mathbb R)$ $\widetilde T_\alpha\vp{T}^\beta$ and $gl(1,\mathbb R)$ $T_0=Y_\alpha Z^\alpha$ ones. Higher-order generators $Y_\alpha Z^{\beta(L)}$ divide into
\beq\label{horder-bos} 
\widetilde T_\alpha\vp{\widetilde T}^{\beta(L)}=Y_\alpha Z^{\beta(L)}-\frac{1}{L+3}(YZ)\de_\alpha^{\beta(1)}Z^{\beta(L-1)} 
\eeq 
and $T_0Z^{\beta(L-1)}$. Expression (\ref{horder-bos}) is an obvious generalization of $\wt T_\alpha\vp{\wt T}^\beta$ from (\ref{gl-string-comp}) to the case $L>1$.

Proceeding to $TSA$ superalgebra, from (\ref{aq-relations}) one infers that the D.B. relations of order $L$ and $M$ generators close on order $L+M-1$ generators. So that order 1 generators, i.e. $gl(4|4,\mathbb R)$ ones (\ref{gl-string-comp}), play a special role: D.B. relations of the generators of an arbitrary order $L$ with those of order 1 yield again order $L$ generators. This feature can be used to characterize irreducible higher-order generators.

Thus the form of irreducible order 2 generators can be found by D.B.-commuting
corresponding bosonic generator (\ref{horder-bos}) with order 1 
supersymmetry generators $Q_i\vp{Q}^\beta$ and $Q_\alpha\vp{Q}^j$,
dividing generators that appear on the r.h.s. into
irreducible $SL(4,\mathbb R)\times SL(4,\mathbb R)$ tensors, then 
D.B.-commuting them with $Q_i\vp{Q}^\beta$ and $Q_\alpha\vp{Q}^j$ 
and so on. In such a way we obtain 
\beq\label{ib-g2d2}
\{Q_i\vp{Q}^\beta(\s),\wt T_{\gamma}\vp{\wt
T}^{\delta(2)}(\s')\}_{D.B.}=\left(\de^\beta_{\gamma}
Q_{i}\vp{Q}^{\delta(2)}-\frac15\de^{(\de_1}_{\gamma}Q_{i}\vp{Q}^{\de_2)\beta}\right)(\s)\de(\s-\s'),
\eeq 
where 
\beq 
Q_{i}\vp{Q}^{\delta(2)}=\eta_i Z^{\de(2)} 
\eeq 
D.B.-commutes with $Q_i\vp{Q}^\beta$. 
Analogously calculation of D.B. relations of $\wt T_{\gamma}\vp{\wt
T}^{\delta(2)}$ and $Q_\alpha\vp{Q}^j$ yields 
\beq\label{aj-gd2} 
\{Q_\alpha\vp{Q}^j(\s),\wt T_{\gamma}\vp{\wt
T}^{\delta(2)}(\s')\}_{D.B.}=-\left(\de_\alpha^{(\de_1}\wt Q_\gamma\vp{\wt Q}^{\de_2)j}-\frac15\de_\gamma^{(\de_1}\wt Q_\alpha\vp{\wt Q}^{\de_2)j}\right)(\s)\de(\s-\s'), 
\eeq 
where another order 2 supersymmetry generator  
\beq 
\wt Q_\gamma\vp{\wt Q}^{\de i}=\wt T_\gamma\vp{\wt T}^{\de}\xi^i 
\eeq 
D.B.-commutes with $Q_\alpha\vp{Q}^j$. 
Applying $Q_\alpha\vp{Q}^j$ to 
$Q_{k}\vp{\wt Q}^{\de(2)}$ gives 
\beq\label{aj-gkd2}
\begin{array}{rl}
\{Q_\alpha\vp{Q}^j(\s),Q_{k}\vp{\wt Q}^{\de(2)}(\s')\}_{D.B.}=&\de^j_k\wt T_{\alpha}\vp{\wt T}^{\de(2)}(\s)\de(\s-\s')\\[0.2cm] 
+&\de_\alpha^{(\de_1}\left(\wt T_k\vp{wt T}^{\de_2)j}+\de_k^j\left(\frac{9}{40}T-\frac{1}{40}U\right)Z^{\de_2)}\right)(\s)\de(\s-\s') 
\end{array}
\eeq 
and similarly 
\beq\label{ib-gdl} 
\begin{array}{rl}
\{Q_i\vp{Q}^\beta(\s),\wt Q_{\gamma}\vp{\wt Q}^{\de l}(\s')\}_{D.B.}=&\left(\de^\beta_\gamma\wt T_i\vp{\wt T}^{\de l}-\frac14\de^\de_\gamma\wt T_i\vp{\wt T}^{\beta l}\right)(\s)\de(\s-\s')\\[0.2cm] 
+&\de_i^l\left[\wt T_\gamma\vp{\wt T}^{\beta\de}+\left(\frac{9}{40}T-\frac{1}{40}U\right)\left(\de^\beta_\gamma Z^\de-\frac14\de^\de_\gamma Z^\beta\right)\right](\s)\de(\s-\s'),
\end{array}
\eeq 
where 
\beq 
\wt T_i\vp{\wt T}^{\beta\! j}=\wt T_i\vp{\wt T}^jZ^\beta.
\eeq
Continuing further one recovers the set of irreducible order 2 generators 
\beq\label{level2-1}
\begin{array}{c}
\wt T_{\alpha}\vp{\wt T}^{\beta(2)},\quad\wt T_{i}\vp{\wt T}^{\alpha j},\quad T_\alpha\vp{T}^{j[2]}=Y_\alpha\xi^{j[2]};\\[0.2cm] 
Q_{i}\vp{Q}^{\alpha(2)},\quad\wt Q_\alpha\vp{\wt Q}^{\beta j},\quad\wt Q_i\vp{\wt Q}^{j[2]}=\eta_i\xi^{j[2]}-\frac13(\eta\xi)\de^{[j_1}_i\xi^{j_2]}
\end{array}
\eeq 
and 
\beq\label{level2-2}
\begin{array}{c}
TZ^\alpha,\quad UZ^\alpha,\quad T\xi^i,\quad U\xi^i.
\end{array}
\eeq
The operators associated with the generators (\ref{level2-2}), as will be shown below, are not the primary fields in the world-sheet CFT and hence corresponding symmetries are broken at the quantum level. Since these generators appear on the r.h.s. of (\ref{aj-gkd2}), (\ref{ib-gdl}) this implies breaking of the order 2 supersymmetries $Q_{i}\vp{\wt Q}^{\alpha(2)}$, $\wt Q_{\alpha}\vp{\wt Q}^{\beta\! j}$ and, in view of (\ref{ib-g2d2}), (\ref{aj-gd2}) breaking of the bosonic symmetry generated by $\wt T_{\gamma}\vp{\wt T}^{\delta(2)}$. So that classical odrer 2 symmetries break in the quantum theory.

For order $L>2$ calculation of D.B. relations of the corresponding bosonic generator (\ref{horder-bos}) and order 1 supersymmetry generators gives 
\beq\label{ib-gd(l)}
\{Q_i\vp{Q}^\beta(\s),\wt T_{\gamma}\vp{\wt
T}^{\delta(L)}(\s')\}_{D.B.}=\left(\de^\beta_\gamma 
Q_{i}\vp{Q}^{\delta(L)}-\frac{1}{L+3}\de^{\de(1)}_{\gamma}Q_{i}\vp{Q}^{\de(L-1)\beta}\right)(\s)\de(\s-\s'),
\eeq 
and 
\beq\label{aj-gd(l)} 
\{Q_\alpha\vp{Q}^j(\s),\wt T_{\gamma}\vp{\wt
T}^{\delta(L)}(\s')\}_{D.B.}=-\left(\de_\alpha^{\de(1)}\wt Q_\gamma\vp{\wt Q}^{\de(L-1)j}-\frac{1}{L+3}\de_\gamma^{\de(1)}\wt Q_\alpha\vp{\wt Q}^{\de(L-1)j}\right)(\s)\de(\s-\s'), 
\eeq 
where order $L$ supersymmetry generators are defined by the expressions 
\beq 
Q_{i}\vp{Q}^{\delta(L)}=\eta_i Z^{\de(L)},\quad\wt Q_\gamma\vp{\wt Q}^{\de(L-1)j}=\wt T_\gamma\vp{\wt T}^{\de(L-1)}\xi^j.
\eeq 
Their D.B. relations with order 1 supersymmetry generators read 
\beq\label{aj-kd(l)}
\begin{array}{rl}
\{Q_\alpha\vp{Q}^j(\s),Q_{k}\vp{Q}^{\de(L)}(\s')\}_{D.B.}=&\de^j_k\wt T_{\alpha}\vp{\wt T}^{\de(L)}(\s)\de(\s-\s')+\de_\alpha^{\de(1)}\left[\wt T_k\vp{wt T}^{\de(L-1)j}\right.\\[0.2cm] 
+&\left.\de_k^j\left(\frac{L+7}{8(L+3)}T-\frac{L-1}{8(L+3)}U\right)Z^{\de(L-1)}\right](\s)\de(\s-\s')
\end{array}
\eeq 
and 
\beq\label{ib-gd(l-1)l} 
\begin{array}{rl}
\{Q_i\vp{Q}^\beta(\s),\wt Q_{\gamma}\vp{\wt Q}^{\de(L-1)l}(\s')\}_{D.B.}=&\left(\de^\beta_\gamma\wt T_i\vp{\wt T}^{\de(L-1)l}-\frac{1}{L+2}\de^{\de(1)}_\gamma\wt T_i\vp{\wt T}^{\beta\de(L-2)l}\right)(\s)\de(\s-\s')\\[0.2cm] 
+&\de_i^l\left[\wt T_\gamma\vp{\wt T}^{\beta\de(L-1)}+\left(\frac{L+7}{8(L+3)}T-\frac{L-1}{8(L+3)}U\right)\right.\\[0.2cm] 
\times &\left.\left(\de^\beta_\gamma Z^{\de(L-1)}-\frac{1}{L+2}\de^{\de(1)}_\gamma Z^\beta Z^{\de(L-2)}\right)\right](\s)\de(\s-\s'), 
\end{array}
\eeq 
where 
\beq 
\wt T_k\vp{wt T}^{\de(L-1)j}=\wt T_k\vp{\wt T}^jZ^{\de(L-1)}. 
\eeq 
Continuing further calculation of D.B. relations of $gl(4|4,\mathbb R)$ supersymmetry generators and order $L$ generators allows to find complete set of irreducible order $L$ bosonic 
\beq\label{orderl-bgen} 
\begin{array}{c}
\wt T_\alpha\vp{\wt T}^{\beta(p)j[q]}=\wt T_\alpha\vp{\wt T}^{\beta(p)}\xi^{j[q]},\ q=0,2,4,\ p+q=L;\\[0.2cm]
\wt T_i\vp{\wt T}^{\beta(p)j[q]}=\wt T_i\vp{\wt T}^{j[q]}Z^{\beta(p)},\ q=1,3,\ p+q=L 
\end{array}
\eeq 
and fermionic generators
\beq\label{orderl-fgen} 
\begin{array}{c}
\wt Q_\alpha\vp{\wt Q}^{\beta(p)j[q]}=\wt T_\alpha\vp{\wt T}^{\beta(p)}\xi^{j[q]},\ q=1,3,\ p+q=L;\\[0.2cm] 
\wt Q_i\vp{\wt Q}^{\beta(p)j[q]}=\wt Q_i\vp{\wt Q}^{j[q]}Z^{\beta(p)},\ q=0,2,4,\ p+q=L. 
\end{array}
\eeq 
Relevant (traceless) products of bosonic components of supertwistors are defined in (\ref{horder-bos}) and the definition of (traceless) products of fermionic components is given in (\ref{gl-string-comp}), (\ref{level2-1}) and by the expressions 
\beq 
\begin{array}{c} 
\wt T_i\vp{\wt T}^{j[3]}=\eta_i\xi^{j[3]}-\frac12(\eta\xi)\de^{[j_1}_i\xi^{j_2}\xi^{j_3]},\\[0.2cm] 
Q_i=\eta_i,\quad Q_i\vp{Q}^{j[4]}=\eta_i\xi^{j[4]}. 
\end{array}
\eeq 
There are also generators of the form 
\beq\label{orderl-non-tensor} 
TZ^{\alpha(p)}\xi^{i[q]},\quad UZ^{\alpha(p)}\xi^{i[q]},\ p\geq0,\ 0\leq q\leq4. 
\eeq 
It is these generators that correspond to non-tensor operators in the world-sheet CFT. They are present on the r.h.s. of (\ref{aj-kd(l)}) and (\ref{ib-gd(l-1)l}) implying breaking of order $L$ symmetries in analogy with those of order 2. 

In Berkovits twistor string theory $GL(1,\mathbb R)$ symmetry is gauged so that generators carrying the factor of $T$ are set to zero. However, $GL_t(1,\mathbb R)$ symmetry, being anomalous, cannot be gauged thus the generators carrying the factor of $U$ cannot be put to zero. So we conclude that for any order $L$ it is not possible to find a set of generators with closed D.B. relations that would correspond to the primary fields. As a result the quantum symmetry of the twistor string reduces to $SL(4|4,\mathbb R)\times SL(4|4,\mathbb R)$ for the sector of closed strings and its diagonal subgroup for the sector of open strings.

\subsubsection{Higher-spin symmetries from the world-sheet CFT perspective}

This subsection we devote to consideration of the twistor part of the left-moving world-sheet CFT justifying the arguments above discussion relied on.
To apply the $2d$ CFT technique to the model (\ref{genBerk}) it is helpful to carry out Wick rotation to Euclidian signature world-sheet
\beq
\tau\rightarrow i\s^2,\ \s\rightarrow\s^1\ \Rightarrow\ \s^+\rightarrow z=\s^1+i\s^2,\ \s^-\rightarrow-\bar z=-(\s^1-i\s^2).
\eeq
The following changes of the world-sheet derivatives
\beq
\partial_+\rightarrow\partial_z=\frac12(\partial_1-i\partial_2)\equiv\partial,\quad\partial_-\rightarrow-\partial_{\bar z}=-\frac12(\partial_1+i\partial_2)\equiv-\bar\partial,
\eeq
$2d$ volume element
\beq
d\tau d\s\rightarrow
id\s^1d\s^2=\frac{i}{2}d^2z,
\eeq
and supertwistor components
\beq
\mathcal
Y_A\rightarrow\mathcal Y_{A(z)},\quad\bar{\mc Y}_A\rightarrow-\bar{\mathcal Y}_{A(\bar z)}, 
\eeq
result in the Euclidean action
\beq\label{genBerkE}
S_E=\int d^2z(\mathcal
Y_A\bar\partial\mathcal Z^A+\bar{\mathcal Y}_A\partial\bar{\mathcal Z}^A).
\eeq
Non-trivial OPE's for the supertwistor
components of the left-moving sector, on which we focus,
\beq
Z^\alpha(z)Y_{\beta}(w)\sim\frac{\de^\alpha_\beta}{z-w},\quad\xi^i(z)\eta_j(w)\sim\frac{\de^i_j}{z-w}
\eeq
in terms of the supertwistors can be written as
\beq
\mc
Z^A(z)\mc Y_B(w)\sim\frac{\de^A_B}{z-w},\quad\mc Y_B(z)\mc
Z^A(w)\sim-\frac{(-)^a\de^A_B}{z-w}.
\eeq

By definition primary fields are characterized by the following general form of the OPE with the world-sheet stress-energy tensor
\beq\label{primf-def}
L(z)\mc O(w)\sim\frac{h}{(z-w)^2}\mc O(w)+\frac{1}{z-w}\,\partial\mc O(w),
\eeq
where $h$ is conformal weight of the primary field.\footnote{It is assumed that composite operators depending on a single argument are normal-ordered but normal ordering signs $:\ :$ will be omitted.} The supertwistor part of the left-moving stress-energy tensor for the twistor string model (\ref{genBerk}) equals
\beq
L_{\mathrm{tw}}(z)=-\mathcal Y_A\partial\mathcal Z^A
\eeq
so that $\mc Y_B$ and $\mc Z^A$ are primary fields of conformal weight 1 and 0 respectively.

From the world-sheet CFT perspective the necessary condition for the considered global symmetries to survive in the quantum theory is that their generators become primary fields, i.e. their OPE's with the stress-energy tensor are anomaly free, in other words, on the r.h.s. of (\ref{primf-def}) there should not appear terms with poles of order higher than two. As we find the generators containing the factor of $T$ or $U$ fail to comply with this requirement.

Using the relations
\beq
Y_\gamma\partial Z^\gamma(z)Y_\beta Z^\alpha(w)\sim\frac{\de^\alpha_\beta}{(z-w)^3}-\frac{1}{(z-w)^2}Y_\beta Z^\alpha(w)-\frac{1}{(z-w)}\,\partial(Y_\beta Z^\alpha)(w)
\eeq
and
\beq
\eta_k\partial\xi^k(z)\eta_j\xi^i(w)\sim-\frac{\de^i_j}{(z-w)^3}-\frac{1}{(z-w)^2}\eta_j\xi^i(w)-\frac{1}{z-w}\,\partial(\eta_j\xi^i)(w),
\eeq
it follows that $sl(4|4,\mathbb R)$ generators $\wt T_\alpha\vp{\wt T}^\beta$, $\wt T_i\vp{\wt T}^j$, $Q_\alpha\vp{Q}^j$, $Q_i\vp{Q}^\beta$  and $T$ are primary fields of unit weight, while $U$ is not \cite{Dolan}
\beq
L_{\mathrm{tw}}(z)U(w)\sim\frac{-8}{(z-w)^3}+\frac{1}{(z-w)^2}U(w)+\frac{1}{z-w}\,\partial U(w).
\eeq

Higher-order generators (\ref{orderl-bgen}), (\ref{orderl-fgen}) also become primary fields of unit weight. While OPE's of the generators (\ref{orderl-non-tensor}) with the stress-energy tensor are anomalous 
\beq 
\begin{array}{rl}
L_{\mathrm{tw}}(z)TZ^{\alpha(p)}\xi^{i[q]}(w)&\sim-\frac{p+q}{(z-w)^3}Z^{\alpha(p)}\xi^{i[q]}(w)+\mc O((z-w)^{-2})\\[0.2cm] 
L_{\mathrm{tw}}(z)UZ^{\alpha(p)}\xi^{i[q]}(w)&\sim-\frac{8+p-q}{(z-w)^3}Z^{\alpha(p)}\xi^{i[q]}(w)+\mc O((z-w)^{-2}).  
\end{array}
\eeq 
In the case $p=q=0$ one recovers discussed above OPE's of $gl(1,\mathbb R)$ and $gl_t(1,\mathbb R)$ generators with the stress-energy tensor. For $p\neq0$, $q\neq0$ anomalous terms do not vanish so that associated symmetries are broken. 
Since generators (\ref{orderl-bgen}), (\ref{orderl-fgen}) are linked with other order $L$ generators by order 1 supersymmetries (cf.~Eqs.~(\ref{ib-gd(l)})-(\ref{ib-gd(l-1)l})) it appears that higher-spin symmetry is broken for arbitrary value of $L$ except for $L=1$, for which quantum-mechanically consistent global symmetry is isomorphic to $SL(4|4,\mathbb R)$.

\setcounter{equation}{0}
\section{Conclusion and discussion}

In this paper we performed the analysis of higher-spin global symmetries
of $D=4$ $N=4$ massless superparticle models in supertwistor formulation
extending the consideration of Ref.~\cite{Townsend}. Discussed infinite-dimensional conformal superalgerbas stemming from the $aq(8|8)$ algebra require further study as they could underly $N=4$ supersymmetric extension of interacting higher-spin theories on $AdS_5$ \cite{Alkalaev-Vasiliev}, \cite{Sezgin} and conformal higher-spin theories on $D=4$ Minkowski space-time \cite{Segal}. 
We have also revealed inifinite-dimensional classical symmetries in the Berkovits twistor string model and its extension with ungauged
$GL(1,\mathbb R)$ symmetry. 
Noether current densities 
associated with these symmetries have been constructed in terms of $PSL(4|4,\mathbb R)$ supertwistors. In the generalized twistor string model the D.B.
relations of the Noether current densities have been shown to form the 
TSA inifinite-dimensional
Lie superalgebra, whose finite-dimensional
subalgebra is spanned by $gl(4|4,\mathbb R)$ generators and the generator of constant shifts of the supertwistor components. The full classical symmetry of the twistor string action is generated by the
direct sum of two copies of TSA superalgebra for the left- and
right-movers that for the open string sector are identified on the boundary. Classical
symmetry of the Berkovits model is described by the subalgebra of TSA obtained by going on the
constraint shell $Y_\alpha Z^\alpha+\eta_i\xi^i\approx0$. Its finite-dimensional subalgebra is spanned by $psl(4|4,\mathbb R)$, 'twisted' $gl_t(1,\mathbb R)$ generators and that of shifts of the supertwistor components. 

The fact that the symmetry of twistor string action is
infinite-dimensional is anticipated due to the symmetry
enhancement in $N=4$ super-Yang-Mills theory at zero coupling
\cite{Sundborg}, \cite{Witten-talk}. One could similarly
anticipate infinite-dimensional symmetry of free $N=4$ conformal
supergravity \cite{FrTs} that is present in the spectrum of
Berkovits twistor string on equal footing with $N=4$
super-Yang-Mills theory. Observed infinite-dimensional symmetry
breaking down to $SL(4|4,\mathbb R)$ at the quantum level also agrees with
the higher-spin symmetry breaking in $N=4$ super-Yang-Mills once
the interactions are switched on \cite{Witten-talk}. Looking at
the symmetry enhancement on the stringy side \cite{Sundborg}, \cite{Sagnotti} in the weak coupling regime of 
gauge/gravity duality our results
seem to support the evidence \cite{BdAM} for the tensionless nature of twistor
strings or rather certain equivalence of the limits of
zero and infinite tension \cite{Bandos2014}. Interesting question
is whether other twistor string models \cite{Abou-Zeid}-\cite{MasonSkinner} are invariant under
higher-spin symmetries.

To conclude let us make a few comments on the twistor string
spectrum. There are three kinds of states in the open string
sector of the Berkovits model \cite{BerkWitt}. Twistor counterpart
of $N=4$ superYang-Mills multiplet is described by the vertex
operator 
\beq 
V_{YM}(z)=j_R(z)F_0^R(\mc Z(z)), 
\eeq 
where $j_R$
($R=1,\ldots,\mathrm{dim}G$) represent currents of unit conformal
weight from the current algebra $G$ that enters the
Lagrangian $\mathscr L_{\mathrm{L(R)-mat}}$ in (\ref{genBerk}) and
$F_0^R(\mc Z)$ is a scalar function on the supertwistor space of
homogeneity degree zero. Other options to construct vertices of
overall conformal weight one and homogeneity degree zero are
\beq\label{confsugravertices} 
V_f(z)=\mc Y_A(z)f^A(\mc Z(z)),\quad
V_g(z)=g_A(\mc Z(z))\partial\mc Z^A 
\eeq 
with the supertwistor
functions $f^A(\mc Z)$ and $g_A(\mc Z)$ having homogeneity degrees
($GL(1,\mathbb R)$ charges) +1 and -1. They satisfy the
constraints $\partial_Af^A=\mc Z^Ag_A=0$ and are defined modulo
the gauge invariances $\de f^A=\mc Z^A\mathfrak f$, $\de
g_A=\partial_A\mathfrak g$ to match upon the twistor transform the
states of $N=4$ conformal supergravity \cite{BerkWitt}. As far as
the open string sector of the model (\ref{genBerk}) is concerned
the vertex operators are formally remain the same as above but the
condition of zero homogeneity degree in supertwistor components is
relaxed so that $F^R_a(\mc Z)$ describes not only $N=4$
superYang-Mills states but also all the doubleton supermultiplets
via the pairs of functions $F_{\pm a}(\mc Z)$ having opposite
homogeneity degrees $+a$ and $-a$. In particular, functions
$F_{\pm2}(\mc Z)$ describe $N=4$ Einstein supergravity multiplet.
It is then natural to take $j_R$  corresponding to some Abelian
algebra. The states of $N=4$ Einstein supergravity also 
reside in conformal supergravity vertices
(\ref{confsugravertices}) with the supertwistor functions
constrained by the ansatze $f^A(\mc
Z)=I^{AB}\partial_B F'_{+2}(\mc Z)$ and $g_A(\mc Z)=F'_{-2}(\mc
Z)I_{AB}\mc Z^B$, where $I^{AB}$, $I_{AB}$ are infinity
supertwistors \cite{Adamo}. Supertwistor functions $f^A(\mc Z)$
and $g_A(\mc Z)$ with other values of $GL(1,\mathbb R)$ charges
correspond to higher-spin counterparts of $N=4$ conformal
supergravity multiplet and deserve further study.
In the Berkovits twistor string model important role is
played by the gauged $GL(1,\mathbb R)$ symmetry that allows to
shift conformal weights of the supertwistor fields and
reproduce scattering amplitudes for various helicity configurations of external particles.
In the ungauged case to be able to study scattering amplitudes of
particles from, for instance, doubleton supermultiplets some
additional variables should be introduced. This could impose further restrictions or lead to the determination of the
structure of yet undetermined matter Lagrangians in (\ref{genBerk}).

\section*{Acknowledgements}

The author is grateful to A.A.~Zheltukhin for stimulating discussions.


\begin{thebibliography}{99}
\bibitem{Witten03}
E.~Witten, Perturbative gauge theory as a string theory in twistor space, Comm. Math. Phys. \textbf{252} (2004) 189, \href{http://arxiv.org/abs/hep-th/0312171}{ArXiV:hep-th/0312171}.
\bibitem{Berkovits}
N.~Berkovits, Alternative string theory in twistor space for $N=4$ super-Yang-Mills theory, Phys. Rev. Lett. \textbf{93} (2004) 011601, \href{http://arxiv.org/abs/hep-th/0402045}{ArXiV:hep-th/0402045}.
\bibitem{BerkWitt}
N.~Berkovits and E.~Witten, Conformal supergravity in twistor string theory, JHEP \textbf{0408} (2004) 009, \href{http://arxiv.org/abs/hep-th/0406051}{ArXiV:hep-th/0406051}.
\bibitem{Abou-Zeid}
M.~Abou-Zeid, C.M.~Hull and L.J.~Mason, Einstein supergravity and new twistor string theories, Comm. Math. Phys. \textbf{282} (2008) 519, \href{http://arxiv.org/abs/hep-th/0606272}{ArXiV:hep-th/0606272}.
\bibitem{Skinner}
D.~Skinner, Twistor strings for $N=8$ supergravity, \href{http://arxiv.org/abs/1301.0868}{ArXiV:1301.0868 [hep-th]}.
\bibitem{MasonSkinner}
L.~Mason and D.~Skinner, Ambitwistor strings and scattering equations, JHEP \textbf{1407} (2014) 048, \href{http://arxiv.org/abs/1311.2564}{ArXiV:1311.2564 [hep-th]}.
\bibitem{DolanGoddard}
L.~Dolan and P.~Goddard, Complete equivalence between gluon tree amplitudes in twistor string theory and in gauge theory, JHEP \textbf{1206} (2012) 030, \href{http://arxiv.org/abs/1111.0950}{ArXiV:1111.0950 [hep-th]}.
\bibitem{Dolan}
J.~Corn, T.~Creutzig and L.~Dolan, Yangian in the twistor string, JHEP \textbf{1010} (2010) 076, \href{http://arxiv.org/abs/1008.0302}{ArXiV:1008.0302 [hep-th]}.
\bibitem{Zarembo}
J.~Minahan and K.~Zarembo, The Bethe-ansatz for $N=4$ super Yang-Mills, JHEP \textbf{0303} (2003) 013, \href{http://arxiv.org/abs/hep-th/0212208}{ArXiV:hep-th/0212208}.
\bibitem{Nappi}
L.~Dolan, C.~Nappi and E.~Witten, A relation between approaches to integrability in superconformal Yang-Mills theory, JHEP \textbf{0310} (2003) 017, \href{http://arxiv.org/abs/hep-th/0308089}{ArXiV:hep-th/0308089}.
\bibitem{Townsend}
P.K.~Townsend, Higher-spin symmetries of the massless (super)particle, Class. Quantum Grav. \textbf{8} (1991) 1231.
\bibitem{Shirafuji}
T.~Shirafuji, Lagrangian mechanics of massless particles with spin, Prog. Theor. Phys. \textbf{70} (1983) 18.
\bibitem{Claus}
P.~Claus, M.~Gunaydin, R.~Kallosh, J.Rahmfeld and Y.~Zunger, Supertwistor as quarks of $SU(2,2|4)$, JHEP \textbf{9905} (1999) 019, \href{http://arxiv.org/abs/hep-th/9905112}{ArXiV:hep-th/9905112}.
\bibitem{Ferber}
A.~Ferber, Supertwistors and conformal supersymmetry, Nucl. Phys. \textbf{B132} (1978) 55.
\bibitem{BL}
I.~Bandos and J.~Lukierski, Tensorial central charges and new superparticle models with fundamental spinor coordinates, Mod. Phys. Lett. \textbf{A14} (1999) 1257, \href{http://arxiv.org/abs/hep-th/9811022}{ArXiV:hep-th/9811022}.
\bibitem{BLS}
I.~Bandos, J.~Lukierski and D.~Sorokin, Superparticle models with tensorial central charges, Phys. Rev. \textbf{D61} (2000) 045002, \href{http://arxiv.org/abs/hep-th/9904109}{ArXiV:hep-th/9904109}.
\bibitem{ZU}
A.A.~Zheltukhin and D.V.~Uvarov, An inverse Penrose limit and supersymmetry enhancement in the presence of tensor central 
charges, JHEP \textbf{0208} (2002) 008, \href{http://arxiv.org/abs/hep-th/0206214}{ArXiV:hep-th/0206214}.
\bibitem{Bandos-ICTP}
I.A.~Bandos, BPS preons and tensionless super-$p$-branes in generalized superspace, Phys. Lett. \textbf{B558} (2003) 197, 
\href{http://arxiv.org/abs/hep-th/0208110}{ArXiV:hep-th/0208110}.
\bibitem{BdAPV}
I.A.~Bandos, J.A.~de Azcarraga, M.~Picon and O.~Varela, $D=11$ superstring model with 30 kappa-symmetries and 30/32 BPS 
states in an extended superspace, Phys. Rev. \textbf{D69} (2004) 085007, \href{http://arxiv.org/abs/hep-th/0307106}{ArXiV:hep-th/0307106}.
\bibitem{BdAM}
I.A.~Bandos, J.A.~de Azcarraga and C.~Miquel-Espanya, Superspace formulations of the (super)twistor string, JHEP \textbf{0607} (2006) 005, \href{http://arxiv.org/abs/hep-th/0604037}{ArXiV:hep-th/0604037}.
\bibitem{FrLin-AP90}
E.S.~Fradkin and V.Ya.~Linetsky, Conformal superalgebras of higher
spins, Ann. Phys. \textbf{198} (1990) 252.
\bibitem{Vasiliev-FP88}
M.~Vasiliev, Extended higher-spin superalgebras and their realizations in terms of quantum operators, Fortschr. Phys. \textbf{36} (1988) 33.
\bibitem{FrVasAP87}
E.S.~Fradkin and M.A.~Vasiliev, Candidate for the role of higher-spin symmetry, Ann. Phys. \textbf{177} (1987) 63.
\bibitem{Vasiliev-PRD01}
M.~Vasiliev, Conformal higher spin symmetries of $4d$ massless supermultiplets and $osp(L,2M)$ invariant equations in generalized (super)space, Phys. Rev. \textbf{66} (2001) 066006, \href{http://arxiv.org/abs/hep-th/0106149}{ArXiV:hep-th/0106149}.
\bibitem{FrLin-NPB91}
E.S.~Fradkin and V.Ya.~Linetsky, Cubic interaction in conformal theory of integer higher-spin fields in four-dimensional space-time, Phys. Lett. \textbf{B231} (1989) 97; Superconformal higher-spin theory in the cubic approximation, Nucl. Phys. \textbf{B350} (1991) 274.
\bibitem{Penrose68}
R.~Penrose, Twistor quantization and curved space-time, Int. J. Theor. Phys. \textbf{1} (1968) 61.
\bibitem{PenR}
R.~Penrose and W.~Rindler, Spinors and Space-time. V2.: Spinor and Twistor Methods in Space-time Geometry, Cambridge Univ. Press, 1986.
\bibitem{Marcus}
M.~Gunaydin and N.Marcus, The spectrum of the $S^5$ compactification of the chiral $N=2$, $D=10$ supergravity and the unitary multiplets of $U(2,2|4)$,  Class. Quantum Grav. \textbf{2} (1985) L11.
\bibitem{GMZ}
M.~Gunaydin, D.~Minic and M.~Zagermann, $4d$ doubleton conformal theories, CPT and IIB string on $AdS_5\times S^5$, Nucl. Phys. \textbf{B534} (1998) 96, Erratum ibid. \textbf{538} (1999) 531, \href{http://arxiv.org/abs/hep-th/9806042}{ArXiV:hep-th/9806042}.
\bibitem{GMZ2}
M.~Gunaydin, D.~Minic and M.~Zagermann, Novel supermultiplets of $SU(2,2|4)$ and the $AdS_5/CFT_4$ duality, Nucl. Phys. \textbf{B544} (1999) 737, \href{http://arxiv.org/abs/hep-th/9810226}{ArXiV:hep-th/9810226}.
\bibitem{Chiodaroli}
M.~Chiodaroli, M.~Gunaydin and R.~Roiban, Superconformal symmetry and maximal supergravity in various dimensions, JHEP \textbf{1203} (2012) 093, \href{http://arxiv.org/abs/1108.3085}{ArXiV:1108.3085 [hep-th]}.
\bibitem{Bars}
I.~Bars and M.~Gunaydin, Unitary representations of non-compact supergroups, Comm. Math. Phys. \textbf{91} (1983) 31.
\bibitem{Alkalaev-Vasiliev}
M.A.~Vasiliev, Cubic interactions of bosonic higher spin fields in $AdS_5$, Nucl. Phys. \textbf{B616} (2001) 106,
\href{http://arxiv.org/abs/hep-th/0106200}{ArXiV:hep-th/0106200}. \\
K.B.~Alkalaev and M.A.~Vasiliev, $N=1$ supersymmetric theory of higher spin gauge fields in $AdS_5$ at the cubic order, 
Nucl. Phys. \textbf{B655} (2003) 57, \href{http://arxiv.org/abs/hep-th/0206068}{ArXiV:hep-th/0206068}.
\bibitem{Sezgin}
E.~Sezgin and P.~Sundell, Towards massless higher spin extension of $D=5$, $N=8$ gauged supergravity, JHEP \textbf{0109} 
(2001) 025, \href{http://arxiv.org/abs/hep-th/0107186}{ArXiV:hep-th/0107186}.
\bibitem{Segal}
A.~Segal, Conformal higher-spin theory, Nucl. Phys. \textbf{664} (2003) 59, \href{http://arxiv.org/abs/hep-th/0207212}{ArXiV:hep-th/0207212}.
\bibitem{Sundborg}
Bo Sundborg, Stringy gravity, interacting tensionless strings and massless higher spins, Nucl. Phys. B, Proc. Suppl. \textbf{102-103} (2001) 113, \href{http://arxiv.org/abs/hep-th/0103247}{ArXiV:hep-th/0103247}.
\bibitem{Witten-talk}
E.~Witten, Space-time reconstruction, \href{http://theory.caltech.edu/jhs60/witten/1.html}{http://theory.caltech.edu/jhs60/witten/1.html}.
\bibitem{FrTs}
E.S.~Fradkin and A.A.~Tseytlin, Conformal supergravity, Phys. Rept. \textbf{119} (1985) 233.
\bibitem{Sagnotti}
A.~Sagnotti and M.~Tsulaia, On higher spins and the tensionless limit of string theory, Nucl. Phys. \textbf{B682} (2004) 83, \href{http://arxiv.org/abs/hep-th/0311257}{ArXiV:hep-th/0311257}.
\bibitem{Bandos2014}
I.~Bandos, Twistor/ambitwistor strings and null-superstrings in space-time of $D=4,10$ and $11$ dimensions, \href{http://arxiv.org/abs/1404.1299}{ArXiV:1404.1299 [hep-th]}.
\bibitem{Adamo}
T.~Adamo and L.~Mason, Einstein supergravity amplitudes from twistor-string theory, Class. Quantum Grav. \textbf{29} (2012) 145010, \href{http://arxiv.org/abs/1203.1026}{ArXiV:1203.1026 [hep-th]}.
\end{thebibliography}
\end{document}